\documentclass[a4paper,11pt]{article}
\usepackage{jheppub}

\usepackage{dsfont}
\usepackage{slashed}
\usepackage{amsmath}
\usepackage{amssymb}

\renewcommand{\d}{\mathrm{d}}
\renewcommand{\k}{\mathbf{k}}
\newcommand{\q}{\mathbf{q}}

\renewcommand{\r}{\mathbf{r}}

\newcommand{\Tr}{\widetilde{\mathrm{Tr}}}

\newcommand{\rk}{\r\cdot\k}

\renewcommand{\P}{{\cal P}}
\renewcommand{\O}[1]{{\cal O}\left(\alpha_\mathrm{s}^{#1}\right)}
\def \als{\alpha_\mathrm{s}}

\title{Renormalization of the cyclic Wilson loop}
\author[1]{Matthias Berwein,}
\author[1]{Nora Brambilla,}
\author[2]{Jacopo Ghiglieri,}
\author[1]{and Antonio Vairo}
\affiliation[1]{Physik-Department, Technische Universit\"{a}t M\"{u}nchen,\\James-Franck-Str. 1, 85748 Garching, Germany}
\affiliation[2]{McGill University, Department of Physics,\\3600 rue University, Montreal QC H3A 2T8, Canada}
\emailAdd{matthias.berwein@mytum.de}
\emailAdd{nora.brambilla@tum.de}
\emailAdd{jacopo.ghiglieri@physics.mcgill.ca}
\emailAdd{antonio.vairo@tum.de}
\preprint{TUM-EFT 30/12}
\abstract{In finite-temperature field theory, the cyclic Wilson loop is defined 
as a rectangular Wilson loop spanning the whole compactified time direction. 
In a generic non-abelian gauge theory, we calculate the perturbative expansion 
of the cyclic Wilson loop up to order $g^4$. At this order and after charge renormalization, 
the cyclic Wilson loop is known to be ultraviolet divergent. 
We show that the divergence is not associated with cusps in the contour but is instead  
due to the contour intersecting itself because of the periodic boundary conditions. 
One consequence of this is that the cyclic Wilson loop mixes under renormalization 
with the correlator of two Polyakov loops. The resulting renormalization equation is tested up to order $g^6$ 
and used to resum the leading logarithms associated with the intersection divergence. 
Implications for lattice studies of this operator, which may be relevant for the phenomenology 
of quarkonium at finite temperature, are discussed.}


\begin{document}

\maketitle

\flushbottom

\section{Introduction}
Wilson loops, ever since their introduction~\cite{Wilson:1974sk}, have
been an important tool in the study of non-abelian gauge theories. 
In particular, rectangular loops with one spatial and one time direction, also 
known as static Wilson loops, have played a crucial role in the understanding of QCD, for 
their Coulomb-like behaviour at short distances is a manifestation of asymptotic freedom 
and their area-law behaviour at large distances of confinement. 
Since static Wilson loops in the large time limit are related to the energy of a static quark-antiquark 
pair~\cite{Susskind:1976pi,Brown:1979ya}, static Wilson loops are also important quantities  
for quarkonium studies~\cite{Brambilla:2004wf,Brambilla:2004jw,Brambilla:2010cs}.

At finite temperature, because quarkonium dissociation provides a probe of the pro\-perties of the QCD medium 
created in heavy-ion collisions~\cite{Matsui:1986dk}, Wilson loops and other observables related to Wilson lines
have been widely used and measured on the lattice (see~\cite{Bazavov:2009us} for a review). 
The periodic boundary conditions that characterize the fields in a thermal field theory 
allow indeed for the definition of other gauge-invariant operators besides the Wilson loop.
An example is the trace of the Polyakov loop, i.e. a Wilson line spanning the entire compactified time axis,
and correlators thereof. Gauge-dependent quantities, such as the trace
of the product of two Polyakov loops, have also been studied.

In this paper, we focus on the \emph{cyclic Wilson loop}, i.e. a rectangular
Wilson loop whose time extent spans the entire Euclidean time axis,
i.e. from $0$ to $1/T$, where $T$ is the temperature.
This quantity can be seen as a gauge-invariant completion of the aforementioned product of two Polyakov loops. 
In particular, we study its renormalization. The motivation for this study comes from Ref.~\cite{Burnier:2009bk}, 
where an order $g^4$ perturbative calculation in dimensional regularization 
showed that the thermal expectation value of the cyclic Wilson loop is ultraviolet (UV) divergent after charge renormalization. 

It is known that a smooth Wilson loop  
is finite after charge renormalization in dimensional regularization~\cite{Polyakov:1980ca,Dotsenko:1979wb}. 
If the contour has cusps, additional UV divergences occur. These divergences are called
\emph{cusp divergences} and the coefficients multiplying them, and hence their anomalous dimensions, 
depend only on the angle at the cusp. Such divergences are renormalized through a multiplicative constant. 
This constant and the associated \emph{cusp anomalous dimension} are known 
to two loops in QCD~\cite{Korchemsky:1987wg} and to three loops in ${\cal N}=4$
supersymmetric Yang--Mills theory~\cite{Correa:2012nk}. 
Furthermore, in~\cite{Brandt:1981kf}, the \emph{intersection divergence}, 
i.e. a UV divergence arising from an otherwise smooth contour intersecting itself, 
was first considered; it was shown that its renormalization is nontrivial and mixes all 
possible loops and correlators of loops sharing the same geometry.

In this paper, we will show that the divergence appearing in the cyclic
loop is an intersection divergence, the intersection being caused by
the periodic boundary conditions.  The renormalization equation mixes the cyclic
Wilson loop with the correlator of two Polyakov loops. This equation holds to all orders.
We will explicitly check this up to order~$g^6$.
Furthermore, the related renormalization group equations allow 
to resum logarithms associated with the intersection divergence; 
we will explicitly provide the resummed expression valid at leading logarithmic accuracy.

The paper is organized as follows. In Sec.~\ref{sec_properties}, we
introduce some properties of Wilson loops and cyclic
Wilson loops that will be relevant for the rest of the paper. 
In Sec.~\ref{sec_pert}, we compute the cyclic loop at order $g^4$ at short
distances $r\ll 1/T$. Our result, UV divergence included, reproduces
the short-distance limit of a calculation first presented in~\cite{Burnier:2009bk}. 
In Sec.~\ref{sec_renorm}, we introduce the renormalization of intersection divergences 
derived in~\cite{Brandt:1981kf} and apply it to the cyclic Wilson loop, obtaining its renormalization equation. 
The perturbative expression of the cyclic Wilson loop is used to determine the order $g^2$
contribution to the renormalization constant appearing in the renormalization equation.
The renormalization-group equation is solved at leading logarithmic accuracy. 
In Sec.~\ref{sec_als3}, we examine the structure of intersection
divergences up to order $g^6$; we show how our equation
correctly cancels these divergences, providing a non-trivial test
of the renormalization equation. In Sec.~\ref{sec_largedistance}, 
we show that the same equation that renormalizes the cyclic Wilson loop 
at short distances renormalizes it at long distances. 
Our expression for the UV divergence at long distance disagrees with 
the corresponding one that can be found in~\cite{Burnier:2009bk}; 
we analyze the origin of the disagreement. Finally, in Sec.~\ref{sec_concl}, we
draw our conclusions, emphasizing the relevance of the result for a proper 
lattice evaluation of the cyclic Wilson loop. 
Some of the results presented here can be found in~\cite{matthias}.

\section{Definition and properties of the cyclic Wilson loop}
\label{sec_properties}
Before we turn to the issue of the renormalization of the cyclic Wilson loop, 
we will start by giving some general properties that will
be instrumental in the following discussion. In Euclidean spacetime, 
the cyclic Wilson loop, $W_c$, is defined as the product of four straight Wilson lines, $U$,  
two of which extend in the time direction from $0$ to $1/T$ 
while the other two stretch in the spatial direction $\r$ from $-\r/2$ to $\r/2$:\footnote{Throughout 
the paper boldface characters will refer to three-vectors and italic characters to their modulus,
e.g.  $r\equiv\vert \r \vert$.} 
\begin{equation}
W_c=\left\langle \Tr\left[
U_0\left(\tfrac{\r}{2},-\tfrac{\r}{2}\right)U_{-\frac{\r}{2}}\left(0,1/T\right)
U_{1/T}\left(-\tfrac{\r}{2},\tfrac{\r}{2}\right)U_{\frac{\r}{2}}\left(1/T,0\right)\right]\right\rangle\,,
\end{equation}
where $\langle O\rangle$ stands for the thermal average of the operator $O$. 
The Wilson lines, $U$, are defined as 
\begin{eqnarray}
U_{\tau}\left(\tfrac{\r}{2},-\tfrac{\r}{2}\right)
&=& \P\exp\left[ig\int_0^1\d s \; \r \cdot {\bf A}\left(\left(s-\tfrac{1}{2}\right)\r,\tau\right)\right],
\nonumber\\
U_{\frac{\r}{2}}\left(\tau_2,\tau_1\right)
&=& \P\exp\left[ig\int_{\tau_1}^{\tau_2} d \tau \; A_0\left(\tfrac{\r}{2},\tau\right)\right],
\label{WLine}
\end{eqnarray}
where $A_\mu=A_\mu^aT^a$ is the gluon field in Euclidean spacetime with boundary condition 
$A_\mu(\r,\tau+1/T) = A_\mu(\r,\tau)$, $\P$ stands for path ordering 
of the colour matrices and $\Tr$ denotes the trace in colour space divided by the number of colours $N_c$.\footnote{
We are working with the generators $T^a$ in $A_\mu=A_\mu^aT^a$ in the fundamental representation of $SU(N_c)$.
}
The four Wilson lines together give a rectangular path with sides $1/T$ and $r$.
If we expand the cyclic Wilson loop in the coupling $g$ and take heed of the path-ordering prescription, we get
\begin{equation}
 W_c=\left\langle\Tr\left[\sum_{n=0}^\infty(ig)^n\int\hspace{-11pt}\Box\hspace{6pt}\d z_{1\, \mu}\;A_\mu\left(z_1\right)
\int^{z_1}\hspace{-20pt}\Box\hspace{10pt}\d z_{2\,\nu}\;A_\nu\left(z_2\right)\cdots
\int^{z_{n-1}}\hspace{-30pt}\Box\hspace{20pt}\d z_{n\,\rho}\;A_\rho\left(z_n\right)\right]\right\rangle\,.
\label{thseries}
\end{equation}
The integral $\displaystyle \int^{z_{i}}\hspace{-19pt}\Box \hspace{6pt}\d z_{i+1\,\mu}$ 
is performed along the rectangular path and ends at the point $z_i$ as a consequence 
of the path-ordering prescription for the colour matrices. Note that 
the gauge fields are instead time ordered~\cite{Brown:1979ya}.
The thermal average \eqref{thseries} is a series of Feynman diagrams 
each of them characterized by $n$ external points $z_i$ lying on the contour of the rectangular path
over which we integrate according to the path-ordering prescription.
Sticking to the terminology found e.g. in Ref.~\cite{Brandt:1982gz}, 
it is convenient to refer to these external points, each one of which 
gets a factor $ig$, as \emph{line vertices}, as opposed to \emph{internal vertices}
connecting gluons to other gluons, ghosts or light quarks. The corresponding Feynman rule for line vertices reads
\begin{figure*}[ht]
 \begin{minipage}{0.5\linewidth}
  \flushright
  \includegraphics[width=0.25\linewidth]{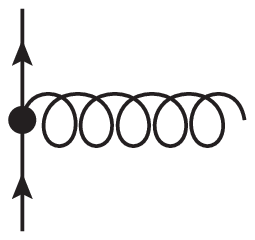}
 \end{minipage}
 \begin{minipage}{0.18\linewidth}
  \vspace{-9pt}
  \begin{equation*}=\,ig\,T^a\int\d z_{\mu}\,.\end{equation*}
 \end{minipage}
 \hfill
 \begin{minipage}{0.1\linewidth}
  \vspace{-8pt}
  \begin{equation}\,\label{linevertex}\end{equation}
 \end{minipage}
\end{figure*}

\begin{figure}[ht]
\centering
\includegraphics[width=0.8\linewidth]{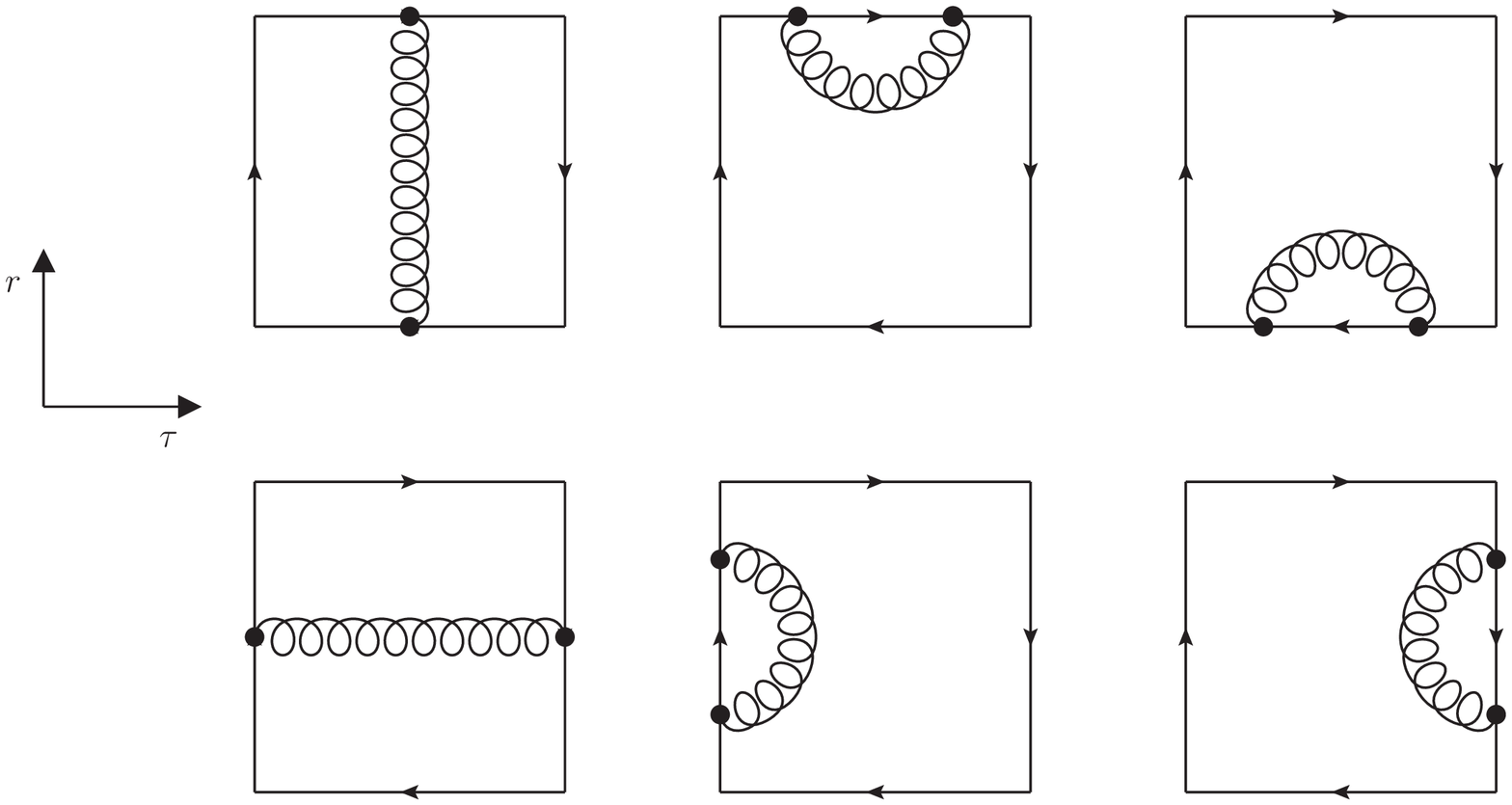}
\caption{Diagrams that can in principle contribute at order $\als$ to the cyclic Wilson loop in gauges where $D_{0i}=0$. 
The time, $\tau$, runs along the horizontal axis and the spatial direction $\r$ along the vertical axis.}
\label{tree}
\end{figure}

Figure~\ref{tree} shows all order $\als$ diagrams that can in principle contribute to the cyclic Wilson loop.\footnote{  
This is the case for gauges in which $D_{0i}=0$, such as the Feynman or
Coulomb gauges. In a generic gauge, there would be also four extra diagrams 
where the gluon connects temporal with spatial lines.}
They represent the first non-trivial contribution in the perturbative series:
\begin{equation}
W_c^{\mathrm{{\cal O}(\als)}}=
(ig)^2C_F\int\hspace{-11pt}\Box\hspace{6pt}\d z_{1\mu}\int^{z_1}\hspace{-20pt}\Box\hspace{10pt}\d z_{2\nu}\,
D_{\mu\nu}\left(z_1,z_2\right)\,,
\end{equation}
where $C_F=\Tr\left[T^aT^a\right]= \left(N_c^2-1\right)/(2N_c)$ is
the quadratic Casimir in the fundamental representation. 

It has been shown in Refs.~\cite{Gatheral:1983cz} and~\cite{Frenkel:1984pz} that the perturbative series 
for a Wilson loop can be rearranged and exponentiated as
\begin{equation}
 W_c=\sum_\gamma C(\gamma)W(\gamma)=\exp\left[\sum_{\gamma\in\mathrm{2PI}}\widetilde{C}(\gamma)W(\gamma)\right]\,,
\label{expth}
\end{equation}
where $W(\gamma)$ stands for the value of a diagram $\gamma$ without
its \emph{colour factor} given by $C(\gamma)$; 
$C(\gamma)$ is the trace divided by $N_c$ of all colour matrices, $T^a$,
contracted with all colour structure constants appearing in the diagram $\gamma$.
Equation  \eqref{expth} states that the sum over all diagrams can be exponentiated 
in such a way that the exponent contains the sum over a subset of these diagrams only, 
called \emph{two-particle irreducible} (2PI) diagrams. However, the colour
factors of the diagrams in the exponent have to be replaced by new factors 
$\widetilde{C}(\gamma)$. These factors are called \emph{colour-connected} 
\cite{Frenkel:1984pz}  or \emph{maximally non-abelian} coefficients \cite{Gatheral:1983cz}. 
We will stick to the former expression and give an explicit definition below.

\begin{figure}[ht]
 \centering
 \includegraphics[width=0.5\linewidth]{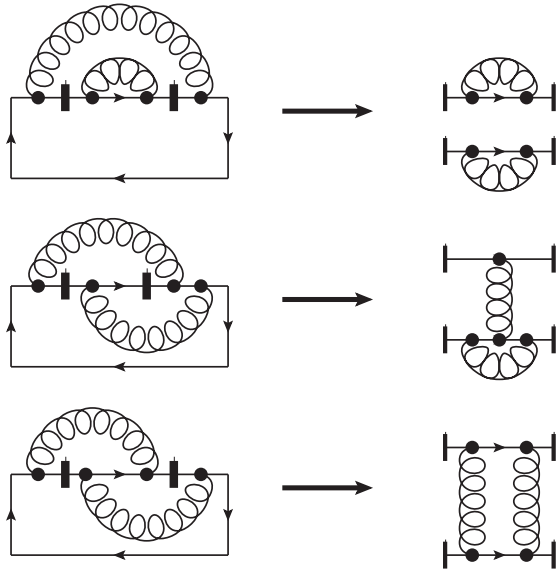}
 \caption{Example of a two-particle reducible (top) and an irreducible diagram (center and bottom).}
 \label{2PI}
\end{figure}

The term two-particle irreducible was coined after the one-particle
irreducible diagrams that occur e.g. in the resummation of the
geometric series of self-energy diagrams for the full propagator
(indeed, Ref.~\cite{Brandt:1982gz} uses the expression
one-particle irreducible, but we will stick to the terminology found
e.g. in Ref.~\cite{Korchemsky:1987wg}). Here ``particle'' refers to
the contour only: if it can be cut in two places in such a way
that the resulting pieces are not connected through gluons, then such
a diagram is reducible. If there is no possible way to cut the contour
twice that leads to two disconnected pieces, then such a diagram is called irreducible.
This is shown in Fig.~\ref{2PI}. The first diagram is reducible,
because if cut as indicated, the two resulting pieces do not exchange
gluons between themselves. These two pieces are given schematically on
the right, dropping all specifics of the contour except for how the
gluons are attached to it. The diagram shown in the second and third line 
cannot be separated like the first diagram, regardless of how it is cut. 
All possibilities reduce to one of the two cases shown.
Of course, if one would cut out a piece without any gluons attached to
it, then every diagram would be reducible making this definition
meaningless, so this trivial case is excluded. Sometimes 2PI diagrams are also
called rainbow irreducible (cf. Ref.~\cite{Korchemsky:1987wg}).

For further use it will be convenient to call a diagram \emph{connected}, if every line vertex is connected
to every other line vertex through gluons, internal vertices and
possibly loops of quarks, gluons and ghosts, otherwise it will be called \emph{unconnected}. 
Connected diagrams are irreducible, while unconnected diagrams,  
like those in Fig.~\ref{2PI}, may or may not be reducible.

\begin{figure}[ht]
 \centering
 \includegraphics[width=\linewidth]{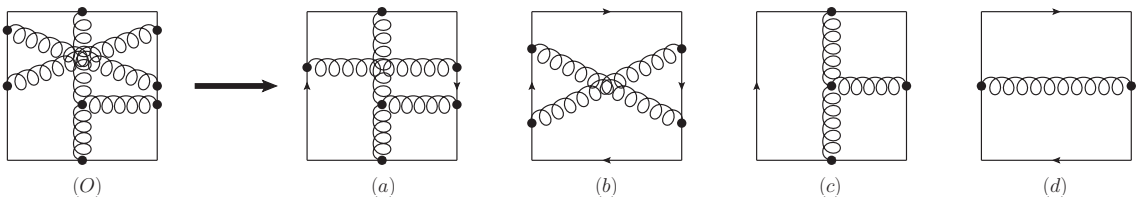}
 \caption{A list of all subdiagrams (called $\gamma_a$ etc.) obtained from the diagram on the left $\gamma_O$.}
 \label{Sub}
\end{figure}

With this terminology we can give a recursive definition
of the colour-connected coefficients. For a connected diagram, the
colour-connected coefficient is equal to its colour factor. We can think
of unconnected diagrams as combinations of connected parts. By combining two diagrams we mean putting their line vertices on the same contour.
A \emph{subdiagram} of
a given diagram can be obtained by removing any number of its connected parts. An example is given in Fig.~\ref{Sub}: the four diagrams to the right represent all subdiagrams of the original diagram $\gamma_O$, obtained by removing one (in $\gamma_a$ and $\gamma_b$) or two (in $\gamma_c$ and $\gamma_d$) of the connected parts.
In general there are several possible sets of subdiagrams
that can be combined to form the original diagram. In our example these are $\{\gamma_a,\gamma_d\}$, $\{\gamma_b,\gamma_c\}$ and $\{\gamma_c,\gamma_d,\gamma_d\}$.
In order to obtain
the colour-connected coefficient for an unconnected diagram we need to
take the products of colour-connected coefficients of the subdiagrams   
for each of these sets and subtract them from the colour factor of the 
original diagram. If there are several ways of combining the subdiagrams
to form the original diagram, then the product of
colour-connected coefficients needs to be multiplied by the number of  
possible combinations.
For example, there are two ways to combine subdiagrams $\gamma_a$ and $\gamma_d$: the left line vertex of $\gamma_d$ can be put above or below the left line vertex of $\gamma_a$; the same applies for the combination of $\{\gamma_c,\gamma_d,\gamma_d\}$.
If a subdiagram appears $n$ times in a set, then
the corresponding product needs to be divided by $n!$. In our example the colour-connected coefficient is
therefore given by
\begin{equation}
 \widetilde{C}(\gamma_O)=C(\gamma_O)-2\,\widetilde{C}(\gamma_a)\widetilde{C}(\gamma_d)-\widetilde{C}(\gamma_b)\widetilde{C}(\gamma_c)-\frac{2}{2!}\,\widetilde{C}(\gamma_c)\left(\widetilde{C}(\gamma_d)\right)^2\,.
\end{equation}
The colour-connected coefficients of unconnected subdiagrams can be obtained by recursive application of this definition. The recursion ends when there are only connected subdiagrams involved. So e.g. for $\gamma_a$ we have
\begin{equation}
 \widetilde{C}(\gamma_a)=C(\gamma_a)-\widetilde{C}(\gamma_c)\widetilde{C}(\gamma_d)=C(\gamma_a)-C(\gamma_c)C(\gamma_d)\,.
\end{equation}
Although the definition does not require subdiagrams to be 2PI, we may neglect reducible subdiagrams in the calculation of colour-connected coefficients, since their colour-connected coefficients are zero.

\begin{figure}[ht]
 \centering
 \includegraphics[width=0.8\linewidth]{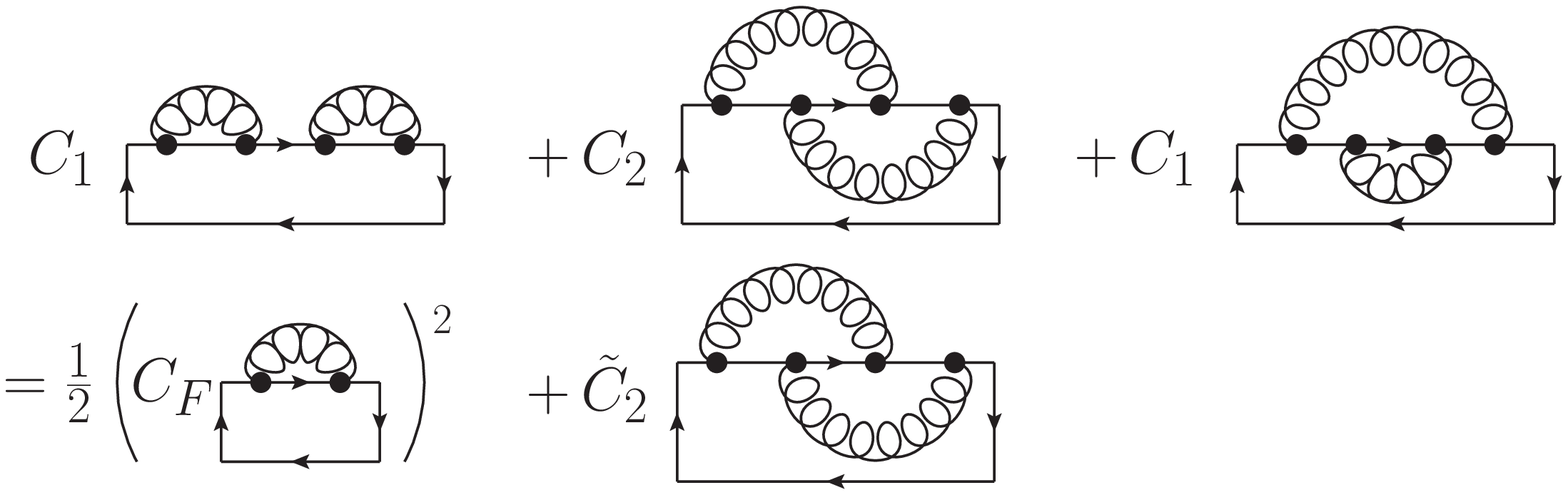}
 \caption{Example of exponentiation; $C_F$ and $\widetilde{C}_2 = -C_FC_A/2$ are colour-connected coefficients.}
 \label{CCC}
\end{figure}

An illustration for Eq.~\eqref{expth} is given in Fig.~\ref{CCC}. 
It shows a series of unconnected diagrams at $\O{2}$ with the colour factors written explicitly 
in front of each diagram. We have
\begin{equation}
C_{1}=\Tr\left[T^aT^aT^bT^b\right]=
C_F^2\hspace{10pt}\mathrm{and}\hspace{10pt}
C_{2}=\Tr\left[T^aT^bT^aT^b\right]=C_F^2-\frac{C_FC_A}{2}\,,
\end{equation}
where $C_A=N_c$ is the quadratic Casimir of the adjoint
representation.  The first and last diagrams are reducible, the one in
the middle is 2PI and its colour-connected coefficient is given by
$\widetilde{C}_2=C_2-2C_F^2/2!=-C_FC_A/2$, since $C_F$ is the
colour-connected coefficient of a one-gluon diagram.  The sum over
the parts proportional to $C_F^2$ of all three diagrams gives the
first term in the second line of Fig.~\ref{CCC}, which we interpret as
the second-order expansion of the exponential of the ${\cal O}(\als)$
diagram, while the remaining 2PI diagram with its modified coefficient
is interpreted as a term from the first-order expansion.

\begin{figure}[th]
 \centering
 \includegraphics[width=0.8\linewidth]{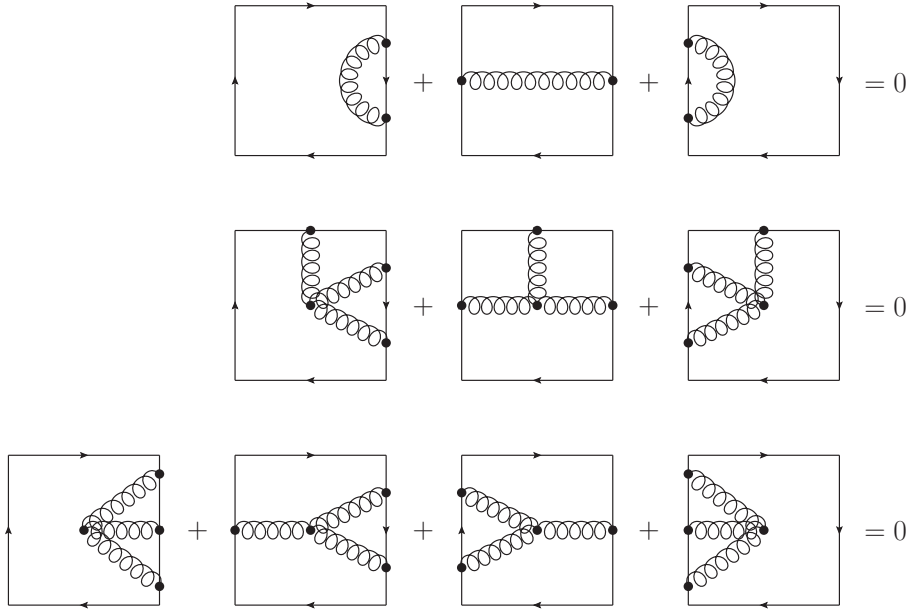}
 \caption{Diagrams canceling because of cyclicity.}
 \label{cyclexamp}
\end{figure}

The exponentiation theorem greatly reduces the number of
diagrams that we need to consider when calculating a Wilson loop, 
for we just need to consider 2PI diagrams.\footnote{We remark that in an abelian theory all colour factors are one 
and only connected diagrams contribute to the exponent of Eq.~\eqref{expth}.} 
For the cyclic Wilson loop, we can neglect yet another group of diagrams. As Eq.~\eqref{WLine} shows,
Wilson lines are unitary operators, whose inverse is given by an
otherwise identical Wilson line but with the direction of the contour
integration reversed. Hence we can rewrite the cyclic Wilson loop as
\begin{equation}
W_c=\left\langle \Tr\left[ U_0\left(\tfrac{\r}{2},-\tfrac{\r}{2}\right)U_{-\frac{\r}{2}}^\dagger\left(1/T,0\right)
U_{1/T}^\dagger\left(\tfrac{\r}{2},-\tfrac{\r}{2}\right)U_{\frac{\r}{2}}\left(1/T,0\right)\right]\right\rangle\,.
\end{equation}
Wilson lines going from $0$ to $1/T$ on the time axis
are Polyakov loop operators. Because they are related to the free energy 
of an infinitely heavy quark in the thermal medium~\cite{McLerran:1981pb}, 
we will refer to them as \emph{quark lines}. The other two Wilson lines will be called
\emph{strings}, because they act as gauge links between the two
Polyakov loop operators. We will use the terms quark line and string 
to denote both the operators and the corresponding contours in spacetime.
Because of the periodic boundary conditions on the Euclidean time, 
the two strings are at the same position and can  
be seen to combine into an adjoint Wilson line. If we expand
one of the quark lines to zeroth order in $\als$ then the strings cancel:
\begin{align}
W_c=&\,\left\langle \Tr\left[
U_0\left(\tfrac{\r}{2},-\tfrac{\r}{2}\right)U_{-\frac{\r}{2}}^\dagger\left(1/T,0\right)
U_{1/T}^\dagger\left(\tfrac{\r}{2},-\tfrac{\r}{2}\right)U_{\frac{\r}{2}}\left(1/T,0\right)\right]
\right\rangle\notag\\
    =&\,\left\langle \Tr\left[
U_0\left(\tfrac{\r}{2},-\tfrac{\r}{2}\right)\left(\mathbb{I} + \dots\right)
U_0^\dagger\left(\tfrac{\r}{2},-\tfrac{\r}{2}\right)U_{\frac{\r}{2}}\left(1/T,0\right)
\right]
\right\rangle\notag\\
    =&\,\left\langle \Tr\left[
U_{\frac{\r}{2}}\left(1/T,0\right)\right]\right\rangle+\dots\,.
\label{cyclicity}
\end{align}
In terms of diagrams, the zeroth-order expansion of a quark line
corresponds to diagrams without any line vertices on that quark
line. Equation~\eqref{cyclicity} shows that in this case all diagrams with
line vertices on the strings cancel against each other and what is left is a 
Polyakov loop. As the second line 
of the equation shows, this cancellation happens
at any order in perturbation theory whenever the colour-singlet component 
of either of the two quark lines is considered.
Some examples are given for illustrative purposes in Fig.~\ref{cyclexamp}.
This leaves only diagrams with all line vertices on the same quark
line or diagrams with some line vertices on both quark lines. 
Line vertices on the strings contribute only in the latter case. Because
this cancellation is directly linked to the periodic boundary 
conditions and to the fact that the cyclic Wilson loop spans the entire compactified time axis,
we will call it \emph{cyclicity cancellation}. 

\section{\texorpdfstring{The cyclic Wilson loop at short distances up to ${\cal O}(\als^2)$}{The cyclic Wilson loop at short distances up to O(alpha\_s\^{}2)}}
\label{sec_pert}
Having illustrated some general properties of the cyclic Wilson loop, 
exponentiation and cyclicity cancellation, we can now turn 
to its calculation up to order $g^4$. In particular, we will investigate  
UV divergences. The analysis of this section will be extended up to 
order $g^6$  in Sec.~\ref{sec_als3}.

The cyclic Wilson loop is a thermal average, therefore, besides 
the scale $1/r$ that characterizes the correlation of 
two quark lines at a distance $r$, it depends on the temperature~$T$.
We will assume that these scales are much larger than the confinement scale, 
so that their contribution may be computed in perturbation theory.
Other scales are also relevant: the in vacuum static energy, which in a weak-coupling 
regime is proportional to $\als/r$ and the screening or Debye mass 
$m_D$, which in a weak-coupling regime is proportional to $gT$.
Our interest is the investigation and ultimate renormalization of the
divergences found in \cite{Burnier:2009bk}, which, as we anticipated,
can be understood and treated as UV intersection divergences. As such,
they are only dependent on the specifics of the contour at the intersection points
 and not on the
details of the thermal scales. In order to be definite, however, we will assume for our 
perturbative calculation
$1/r\gg T\gg m_D \gg \als/r$. A reason for this choice is that
 the same hierarchy was assumed 
for the correlator of two Polyakov loops in~\cite{Brambilla:2010xn} 
and we will need the expression of the correlator of two Polyakov loops for 
renormalizing the cyclic Wilson loop at order $g^6$. Furthermore, comparisons 
with the finite parts of the result of \cite{Burnier:2009bk} are possible 
in this setup. In Section~\ref{sec_largedistance} we will briefly analyze the 
divergent structure at large distances, $rm_D\sim 1$.

The cyclic Wilson loop is gauge invariant, so the choice of the gauge in which  
to perform the calculation is only a matter of convenience. A convenient gauge is the Coulomb gauge 
that we will adopt in this section if not stated otherwise, whereas the analysis 
of Sec.~\ref{sec_als3} will not rely on a specific gauge choice.

The ${\cal O}(\als)$ diagrams are shown in Fig.~\ref{tree}. 
The diagrams in the second row cancel because of
cyclicity (see also Fig.~\ref{cyclexamp}). In the first row, only the
diagram on the left, where the gluon connects to both quark lines,
contributes to the cyclic Wilson loop.
The other two diagrams vanish in dimensional regularization. 
The reason is that the contour integration along the time axis, 
i.e. $\displaystyle \int_0^{1/T}\d\tau \, e^{i\omega_n\tau}$ with 
$\omega_n=2\pi n T$ the  bosonic Matsubara frequencies, selects the zero mode $n=0$, 
but then the remaining integration over the spatial momentum $\k$ depends neither 
on $r$ nor on $T$ and vanishes in dimensional regularization for being scaleless.
Contributions from the scale $m_D$ are of higher order.
This line of argument is gauge independent, so the fact that the only contributing 
diagram is the first diagram on the left in Fig.~\ref{tree} is a gauge-independent 
statement. The diagram gives
\begin{equation}
\ln W_c=\frac{C_F\als}{rT}+\O{2}\,.
\end{equation}

\begin{figure}[ht]
 \centering
 \includegraphics[width=\linewidth]{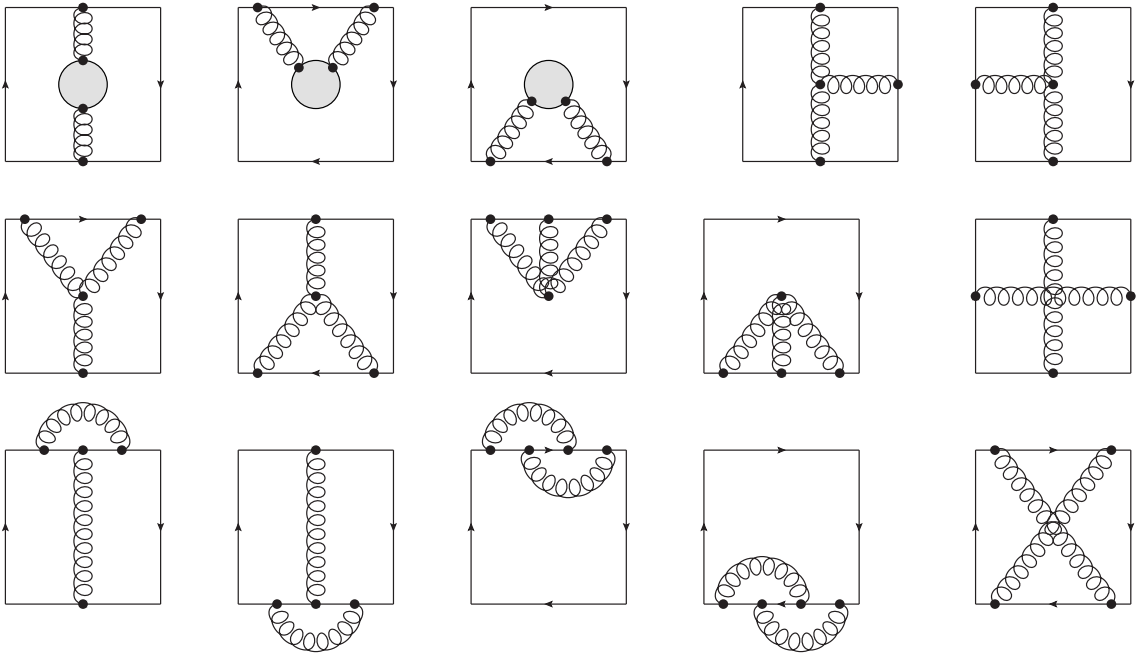}
 \caption{All relevant diagrams at $\O{2}$. As in Fig.~\ref{tree} we
restrict ourselves to gauges where $D_{0i}=0$.
}
 \label{calc}
\end{figure}

At $\O{2}$ all irreducible diagrams that do not cancel through
cyclicity are shown in Fig.~\ref{calc}. The first three diagrams in 
the first row involve the gluon self-energy. This can be split up
into a thermal and a vacuum part, where the thermal part is defined as
the one depending on the Bose--Einstein or Fermi--Dirac distribution functions. 
In Coulomb gauge, the thermal part of the gluon self-energy at zero Matsubara frequency reads
\begin{align}
 \Pi_{00}^{(T)}(0,\k)=&\,\frac{4\als C_A}{\pi}\int_0^\infty\d q\,q\,n_\mathrm{B}(q)
\left[1-\frac{k^2}{2q^2}+\left(\frac{q}{k}-\frac{k}{2q}+\frac{k^3}{8q^3}\right)
\ln\left|\frac{2q+k}{2q-k}\right|\right]\notag
\\
&+\frac{4\als n_f}{\pi}\int_0^\infty\d q\,q\,n_\mathrm{F}(q)
\left[1+\left(\frac{q}{k}-\frac{k}{4q}\right)\ln\left|\frac{2q+k}{2q-k}\right|\right]\,,
\label{tvacpol}
\end{align}
where $n_f$ is the number of massless fermions, 
and $n_\mathrm{B}(q)=1/(\exp(q/T)-1)$ and $n_\mathrm{F}(q)=1/(\exp(q/T)+1)$ are the Bose--Einstein and
Fermi--Dirac distributions respectively. We have taken the zero Matsubara
frequency because, as before, this is the only contribution 
that survives the time integration. The gauge contribution to \eqref{tvacpol} 
can be read from~\cite{Heinz:1986kz}, whereas the fermionic contribution
can be found in textbooks such as~\cite{Kapusta:2006pm}. 
The expression of $\Pi_{00}^{(T)}(0,\k)$ clearly shows that the thermal part of the self-energy 
is not UV divergent. Instead, all three diagrams that involve the gluon self-energy
are IR divergent, but the IR divergences cancel each other~\cite{Brambilla:2008cx}.
Regarding the vacuum part of the gluon-self energy, its expression in Coulomb gauge 
is known and can be read, for instance, from~\cite{Andrasi:2003zf}. 
Summing together vacuum and thermal part, we get from the first three diagrams of Fig.~\ref{calc}
\begin{align}
 &\frac{C_F\als^2}{T} \int\frac{\d^dk}{(2\pi)^d}\frac{e^{i {\bf r}\cdot {\bf k}}}{\k^2}
\left[\left(\frac{31}{9}C_A-\frac{10}{9}n_f\right)
+
\beta_0 \left(\frac{1}{\varepsilon} + \ln 4\pi - \gamma_E - \ln \k^2\right)
\right]
\notag\\
&+\frac{4\pi C_F\als}{T}
\int\frac{\d^3k}{(2\pi)^3}
\left( e^{i {\bf r}\cdot {\bf k}} - 1 \right) 
\left[\frac{1}{\k^2+\Pi_{00}^{(T)}(0,\k)}-\frac{1}{\k^2}\right]\,,
\end{align}
where $\beta_0 = 11 C_A/3 - 2 n_f/3 $ and the first integral has been regularized in $d=3-2\varepsilon$ dimensions.
The integral $\displaystyle \int\frac{\d^dk}{(2\pi)^d}\frac{e^{i {\bf r}\cdot {\bf k}}}{\k^2}$ gives 
$\displaystyle \frac{1}{4\pi r}+{\cal O}(\varepsilon)$, but, 
since the  ${\cal O}(\varepsilon)$ term multiplies a $1/\varepsilon$ pole, we leave it uncomputed
for the time being. The first line is the vacuum part and reproduces the well-known one-loop
contribution to the $Q\overline{Q}$ static potential~\cite{Fischler:1977yf,Billoire:1979ih}. 
The UV divergence in the vacuum part can be removed by charge renormalization.
In the second line, the term in square brackets
simplifies to $-\Pi_{00}^{(T)}(0,\k)/\k^4$ when $k$ is integrated over the momentum 
regions $k\sim 1/r$ and $k\sim T$, whereas the part of $\Pi_{00}^{(T)}(0,\k)$ of order 
$m_D^2$ has to be kept unexpanded when integrating over the momentum region $k\sim m_D$.

The other two diagrams in the first row of Fig.~\ref{calc} are finite; each contributes $C_FC_A\als^2/2$. 
The first four diagrams in the second row give zero, because in
gauges where time and space components do not mix, the three-gluon
vertex for three longitudinal gluons vanishes.  Also all five diagrams
in the third row vanish in Coulomb gauge: the first four, because 
they involve scaleless integrals, and the last one because it is 
proportional to a function with vanishing support along the time axis.

The last diagram in the second row, however, does give a contribution,
which is ultraviolet divergent even after charge renormalization. 
The divergence comes from the vacuum part of the gluon connecting the strings. 
The thermal part is finite and will be given as a series expansion in $rT$, 
which, according to our adopted hierarchy of energy scales, is a small parameter. 
The radius of convergence for this expansion is $rT\leq 1$. We obtain
\begin{equation}
\frac{4C_FC_A\als^2}{T}\int\frac{\d^dk}{(2\pi)^d}
\frac{e^{i\rk}}{\k^2}\left(\frac{1}{\varepsilon}+1+\gamma_E+\ln\pi+\ln r^2\right)
+\frac{2C_FC_A\als^2}{\pi}\sum_{n=1}^\infty\frac{(-1)^n\zeta(2n)}{n(4n^2-1)}(rT)^{2n-1}.
\label{uvdiv}
\end{equation}

The previous calculation shows the specific advantages of the Coulomb gauge:
UV divergences related to charge renormalization occur only in diagrams 
with gluon self-energy insertions, the uncancelled extra divergence 
arises only from the last diagram in the second row of Fig.~\ref{calc}, 
and several diagrams vanish. For comparison, in Feynman gauge only the diagrams 
with a three-gluon vertex in the second row vanish, while the two diagrams on the left
of the third row are also divergent and contribute to both the divergence that is removed 
by charge renormalization and the one which remains after that.
Since the cyclic Wilson loop is gauge invariant, the complete expression up to ${\cal O}(\als^2)$
is the same for both gauges (we have explicitly checked this) and reads
\begin{eqnarray}
\ln W_c&=&\,\frac{C_F\als}{rT} 
\notag\\
&&+ \frac{C_F \als^2}{T} \int\frac{\d^dk}{(2\pi)^d}\frac{e^{i {\bf r}\cdot {\bf k}}}{\k^2}
\left[\left(\frac{31}{9}C_A-\frac{10}{9}n_f\right)
+\beta_0\left(\frac{1}{\varepsilon} + \ln 4\pi - \gamma_E - \ln \k^2 \right)\right]
\notag\\
&&+\frac{4\pi C_F\als}{T}
\int\frac{\d^3k}{(2\pi)^3}
\left(  e^{i {\bf r}\cdot {\bf k}} - 1 \right) \left[\frac{1}{\k^2+\Pi_{00}^{(T)}(0,\k)}-\frac{1}{\k^2}\right]
\notag\\
&&+\frac{2C_FC_A\als^2}{\pi}\sum_{n=1}^\infty\frac{(-1)^n\zeta(2n)}{n(4n^2-1)}(rT)^{2n-1}+C_FC_A\als^2
\notag\\
&&+\frac{4C_FC_A\als^2}{T}
\int\frac{\d^dk}{(2\pi)^d}
\frac{e^{i\rk}}{\k^2}\left(\frac{1}{\varepsilon}+1+\gamma_E+\ln\pi+\ln r^2\right)+\mathcal{O}\left(g^5\right)\,,
\label{finalunrenormalized}
\end{eqnarray}
with $\Pi_{00}^{(T)}(0,\k)$ given by Eq.~\eqref{tvacpol}. 
Equation \eqref{finalunrenormalized} agrees with the
short-distance limit of the result that can be found in Ref.~\cite{Burnier:2009bk};
it thereby confirms their finding that the cyclic Wilson loop is not
finite after charge renormalization.

The divergence in the second line of \eqref{finalunrenormalized} can be removed by charge renormalization.
The divergence coming from the last diagram in the second row of Fig.~\ref{calc}, 
whose analytical expression is in the $1/\varepsilon$ pole in the last line of \eqref{finalunrenormalized}, 
is very peculiar in that it is not of the form of a cusp divergence typically associated to a non-smooth Wilson loop. 
We recall that cusp divergences only depend on the angle $\gamma$ at the cusp. 
In Euclidean spacetime, the leading-order divergence reads in dimensional
regularization\footnote{The angular dependence in a cut-off 
regularization was derived in~\cite{Polyakov:1980ca}.}~\cite{Korchemsky:1987wg}
\begin{equation}
\frac{\als C_F}{2\pi\varepsilon}\left[1+(\pi-\gamma)\cot\gamma\right]\,.
\label{cuspdiv}
\end{equation}
For four right angles,~ like in the case of a rectangular Wilson loop,~ 
Eq.~\eqref{cuspdiv} gives $2\als C_F/(\pi\varepsilon)$; this divergence, being independent 
on $r$ and $T$, can be removed by a multiplicative factor.
This is not the case for the divergence we are discussing here, 
which is of order $\als^2$ and depends on the distance and on the temperature.
In the next section, we will show how to properly renormalize this divergence.

\section{Renormalization}
\label{sec_renorm}
\begin{figure}[ht]
 \centering
 \includegraphics[width=0.8\linewidth]{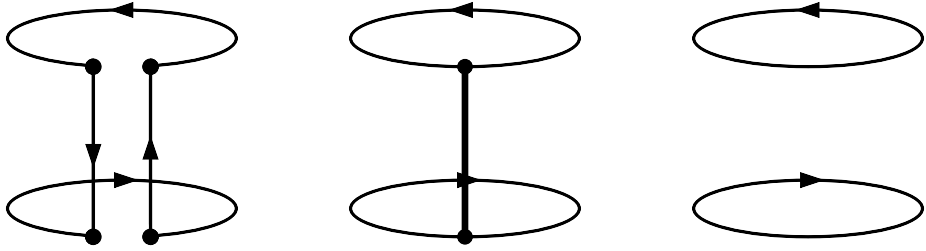}
 \caption{The picture shows the contours of a non-cyclic (left) and a cyclic Wilson loop (middle). 
One can see how the cusp points turn into intersection points. 
The contour of the correlator of two Polyakov loops is shown on the right.}
 \label{cylindric}
\end{figure}

In this section we will assume that
charge renormalization has been carried out, so that we only
need to concern ourselves with the remaining UV divergence in Eq.~\eqref{finalunrenormalized}. 
This divergence is related to the specifics of the contour.
Due to the periodic boundary conditions, the contour of a 
cyclic Wilson loop has support on a Euclidean cylindric
space with $d$ space dimensions extending from $-\infty$ to 
$+\infty$  and one compactified time dimension. 
Because the cyclic Wilson loop wraps around the full time dimension, 
the two strings run actually along the same line.  
This implies that the cyclic Wilson loop has no cusps, but
only intersections, as shown in Fig.~\ref{cylindric}.

It has been shown in Ref.~\cite{Brandt:1981kf} that the expectation
values of Wilson loops with intersections cannot be renormalized by a single
multiplicative constant. One has to consider instead sets of associated
loops and \emph{loop correlators} that mix under renormalization. 
By loop correlators we mean the expectation values (vacuum or thermal) of
products of individually traced loops. These sets of loops and
correlators consist of all possible path-ordering prescriptions for
contours that occupy the same points in spacetime and retain the same
direction everywhere except at the intersection points. 

\begin{figure}[ht]
 \centering
 \includegraphics[width=0.8\linewidth]{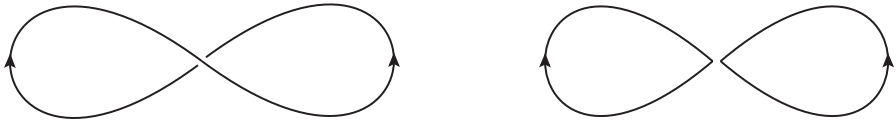}
 \caption{The two possible path orderings at the intersection point 
for a loop with one intersection point. In the figure on the right, the two loops are understood
to be touching at the cusp points; the separation has been introduced
to make clearer that one is dealing with two separate contours.}
 \label{intersection}
\end{figure}

As an illustration consider the simple case shown in Fig.~\ref{intersection}, which 
consists of a smooth curve intersecting itself once at a single point.
Following the curve up to the intersection point, one can either go straight ahead 
thus following the rest of the contour (left picture in Fig.~\ref{intersection}), 
or make a turn onto the way one has come from, thus splitting the contour into two separate loops
(right picture in Fig.~\ref{intersection}). 
Each of the two loops in the right picture, taken on its own, would have a
cusp and be renormalizable through a multiplicative constant. 
However, when considering the product of the loops, there are new divergences 
coming from diagrams with gluon exchanges between the two loops. 
These new divergences are renormalized together with the smooth loop in the left picture, 
for which similar divergences arise at the intersection. 
More precisely, there exist linear combinations of the two loops shown in 
Fig.~\ref{intersection} that are finite and involve coefficients depending only on the 
angle at the intersection point.

In general, a Wilson loop may cross an intersection point several times and
the angles at which the different lines enter the intersection point 
may all be different. In that case, the set of all associated loops is
renormalized by a matrix of renormalization constants, which depend
only upon the angles at the intersection point. When a loop has more
than one intersection point, then the set of associated loops takes on
a tensor-like structure with a renormalization matrix for each
intersection point. If there are additional cusps present, then those
can be taken care of by multiplicative constants. So the general 
formula for the renormalized loops is\footnote{The generalization to
loop functions with more intersections performed in [14] relies on the
assumption that the divergence structure at an intersection point is
completely determined by the local characteristics of the contour at
this point. This is certainly the case when for each intersection
point there are at most two Wilson lines connecting it to other
intersection points. For the cyclic Wilson loop we may therefore use
Eq. (4.1), although it may not be applicable for more general loop
functions.}
\begin{equation}
 W^{(R)}_{i_1 i_2 \dots i_r}=Z_{i_1j_1}(\theta_1)Z_{i_2j_2}(\theta_2)\cdots Z_{i_rj_r}(\theta_r)
Z(\varphi_1)Z(\varphi_2)\cdots Z(\varphi_s)W_{j_1 j_2 \dots j_r}\,.
\label{renormgeneral}
\end{equation}
Here the indices $i_k$ and $j_k$ label the different possible path-ordering prescriptions at the $r$ intersection points, $\theta_k$
denote the sets of angles at those intersection points and $\varphi_l$
stand for the cusp angles at the $s$ additional points of
non differentiability. The loop functions, $W_{j_1 j_2 \dots j_r}$, as well as the renormalized
ones, $W^{(R)}_{i_1 i_2 \dots i_r}$, are defined such that there is a colour trace over each closed Wilson line, 
and each trace is normalized by the number of colours. The trace over closed Wilson loops  
ensures that all loop functions are gauge invariant.
The coupling in $W^{(R)}_{i_1 i_2 \dots i_r}$ is the renormalized coupling.
The matrices $Z$ are the renormalization matrices. 
They are one at leading order in perturbation theory, while the 
explicit expression at higher orders depends on the adopted subtraction scheme.
We will adopt the $\overline{\mathrm{MS}}$ scheme.

\begin{figure}[ht]
 \centering
 \includegraphics[width=0.8\linewidth]{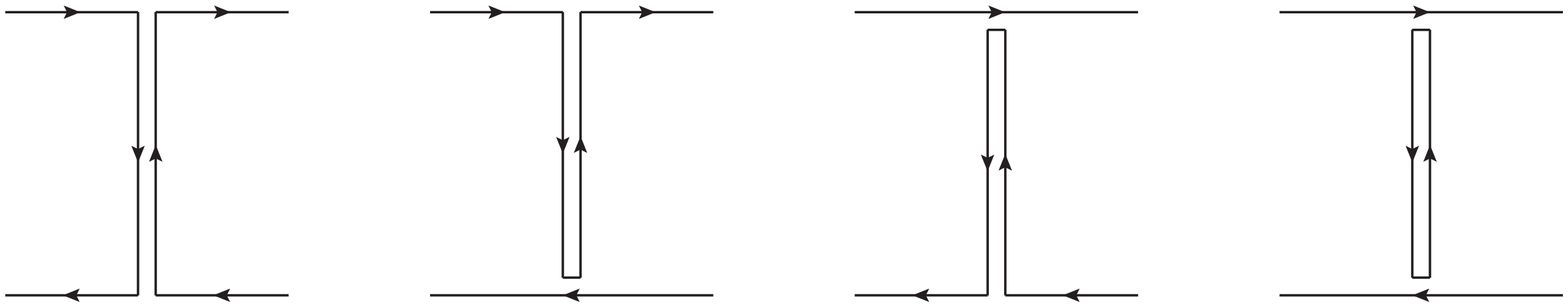}
 \caption{The four different path orderings of the contour of the cyclic Wilson loop that 
correspond from left to right to the loop functions $W_{00}$, $W_{01}$, $W_{10}$ and $W_{11}$ respectively.
The strings are represented by the middle lines.} 
 \label{contours}
\end{figure}

Now we want to apply the results of Ref.~\cite{Brandt:1981kf}, which we have summarized above,
and specifically Eq.~\eqref{renormgeneral} to the case of the cyclic Wilson loop. 
Although it may seem that the cyclic Wilson loop has a continuously infinite number of intersection points, 
namely all the points on the overlapping strings (see the second picture in Fig.~\ref{cylindric}),  
we need to care only about the two endpoints, for the Wilson loop contour does not lead to divergences 
in the other ones.   
As a consequence, we have four possible path orderings (two for each endpoint) when we
consider the possible prescriptions at the intersection points, see  Fig.~\ref{contours}.
We will label the corresponding loop functions $W_{ij}$, where $i$ and $j$ can assume the values 
0 or 1. These are explicitly given by 
\begin{eqnarray}
W_{00}&=&\left\langle\Tr\left[U_0\left(\tfrac{\r}{2},-\tfrac{\r}{2}\right)U_{-\frac{\r}{2}}^\dagger(1/T,0)
U_0^\dagger\left(\tfrac{\r}{2},-\tfrac{\r}{2}\right)U_{\frac{\r}{2}}(1/T,0)\right]\right\rangle= W_c \,,
\notag\\
W_{01}&=&\left\langle\Tr\left[U_0\left(\tfrac{\r}{2},-\tfrac{\r}{2}\right)U_0^\dagger\left(\tfrac{\r}{2},-\tfrac{\r}{2}\right)
U_{\frac{\r}{2}}(1/T,0)\right]\Tr\left[U_{-\frac{\r}{2}}^\dagger(1/T,0)\right]\right\rangle
\notag\\
&=&\left\langle\Tr\left[U_{\frac{\r}{2}}(1/T,0)\right]\Tr\left[U_{-\frac{\r}{2}}^\dagger(1/T,0)\right]\right\rangle \equiv P_c\,,
\notag\\
W_{10}&=&\left\langle\Tr\left[U_0\left(\tfrac{\r}{2},-\tfrac{\r}{2}\right)U_{-\frac{\r}{2}}^\dagger(1/T,0)
U_0^\dagger\left(\tfrac{\r}{2},-\tfrac{\r}{2}\right)\right]\Tr\left[U_{\frac{\r}{2}}(1/T,0)\right]\right\rangle= P_c \,,
\notag\\
W_{11}&=&\left\langle\Tr\left[U_{-\frac{\r}{2}}^\dagger(1/T,0)\right]\Tr\left[U_0\left(\tfrac{\r}{2},-\tfrac{\r}{2}\right)
U_0^\dagger\left(\tfrac{\r}{2},-\tfrac{\r}{2}\right)\right]\Tr\left[U_{\frac{\r}{2}}(1/T,0)\right]\right\rangle = P_c \,.
\end{eqnarray}
We see that the four options give rise to two independent loop functions:
the cyclic Wilson loop, $W_c=W_{00}$, and the correlator of two Polyakov loops 
separated by a distance $\r$, $P_c=W_{01}=W_{10}=W_{11}$.
We will call $P_c$ simply the \emph{Polyakov loop correlator}; 
its contour is shown in the rightmost picture of Fig.~\ref{cylindric}.

We can represent $W_{ij}$ by a four component vector $(W_{00},W_{01},W_{10},W_{11})$ 
that gets renormalized by the tensor product of the two renormalization
matrices corresponding to the two endpoints.
Since the angles at both ends of the string are equal, also
the renormalization matrices are equal. The renormalization equation reads 
\begin{equation}
 \left(
	\begin{array}{c} W_c^{(R)}\\ P_c^{(R)}\\ P_c^{(R)}\\ P_c^{(R)}\end{array}
 \right)=
 \left(
	\begin{array}{c|c}
		Z_{00}\left(
			\begin{array}{cc} Z_{00} & Z_{01}\\ Z_{10} & Z_{11}\end{array}
		\right) &
		Z_{01}\left(
			\begin{array}{cc} Z_{00} & Z_{01}\\ Z_{10} & Z_{11}\end{array}
		\right) \\ \hline
		Z_{10}\left(
			\begin{array}{cc} Z_{00} & Z_{01}\\ Z_{10} & Z_{11}\end{array}
		\right) &
		Z_{11}\left(
			\begin{array}{cc} Z_{00} & Z_{01}\\ Z_{10} & Z_{11}\end{array}
		\right)
	\end{array}
 \right)\left(
	\begin{array}{c} W_c\\ P_c\\ P_c\\ P_c\end{array}
 \right)\,.
\end{equation}
Since the Polyakov loop correlator is finite, having neither
cusps nor intersections, it holds that $P_c^{(R)}=P_c$. 
From this it follows that $Z_{10}$ has to
be zero, otherwise $P_c^{(R)}$ would depend on $W_c$:
\begin{equation}
 \left(
	\begin{array}{c} W_c^{(R)}\\ P_c\\ P_c\\ P_c\end{array}
 \right)=
 \left(
	\begin{array}{cccc}
		Z_{00}^2 & Z_{00}Z_{01} & Z_{00}Z_{01} & Z_{01}^2\\
		0 & Z_{00}Z_{11} & 0 & Z_{01}Z_{11}\\
		0 & 0 & Z_{00}Z_{11} & Z_{01}Z_{11}\\
		0 & 0 & 0 & Z_{11}^2
	\end{array}
 \right)\left(
	\begin{array}{c} W_c\\ P_c\\ P_c\\ P_c\end{array}
 \right)\,.
\end{equation}
The three equations involving $P_c$ furthermore require $Z_{11}=1$ 
and $Z_{01}=1-Z_{00}$, which leaves only one independent renormalization
constant $Z\equiv Z_{00}^2$. So the renormalization condition for the
cyclic Wilson loop and the Polyakov loop correlator reads 
\begin{equation}
 \left(\begin{array}{c} W_c^{(R)} \\ P_c \end{array}\right)=
\left(\begin{array}{cc} Z & (1-Z) \\ 0 & 1 \end{array}\right)
\left(\begin{array}{c} W_c \\ P_c \end{array}\right)\,.
\label{renmatrix}
\end{equation}

We will now determine $Z$.\footnote{
Note that the renormalization constant $Z$ is gauge independent for it relates gauge independent quantities .
} 
For the purpose of determining $Z$ at ${\cal O}(\als)$, we just need to know that $P_c  = 1 \, + {\cal O}(g^3)$.
However, for the forthcoming analysis of Sec.~\ref{sec_als3}, we will need $P_c$ at ${\cal O}(\als^2)$, 
which we give here. The expectation value of the Polyakov loop correlator 
is equal to the square of a single Polyakov loop, $P_L$, plus diagrams 
involving gluon exchanges between both loops. 
The one-gluon exchange diagram vanishes, because it is proportional to $(\mathrm{Tr}[T^a])^2=0$, 
so the first contribution comes from the exchange of two gluons~\cite{McLerran:1981pb,Gross:1980br}.  
The result at $\O{2}$ for the same hierarchy of energy scales that we are assuming here, 
i.e. $1/r\gg T\gg m_D\gg \als/r$, can be read from~\cite{Brambilla:2010xn} 
or extracted in the appropriate limit from ~\cite{McLerran:1981pb,Gross:1980br,Nadkarni:1986cz}. 
The Polyakov loop $P_L= 1+ \delta P_L$ has been computed in~\cite{Burnier:2009bk,Brambilla:2010xn}. 
The full $\O{2}$ result reads
\begin{eqnarray}
P_c &=& 1-\left(C_F^2-\frac{1}{2}C_FC_A\right)\frac{\als^2}{2r^2T^2}+2\,\delta P_L\,,
\label{PL1} \\
\delta P_L &=& \frac{C_F\als m_D}{2T}+C_F\als^2\left[C_A\left(\frac{1}{4}+\ln\frac{m_D}{T}\right)-\frac{n_f}{2}\ln2\right]\,,
\end{eqnarray}
where $\displaystyle m_D^2=\frac{g^2T^2}{3}\left(C_A+\frac{n_f}{2}\right)$. 

Let us now apply the renormalization equation \eqref{renmatrix} to the unrenormalized 
result \eqref{finalunrenormalized}. It is convenient to expand $Z$ in powers of $\als$
(understood as the $2\varepsilon$-dimensional coupling of dimensional regularization):
\begin{equation}
Z=1+Z_1\als\mu^{-2\varepsilon}+Z_2\left(\als\mu^{-2\varepsilon}\right)^2+ {\cal O}(\als^3)\,,
\label{zexpand}
\end{equation}
where $\mu$ is the scale of dimensional regularization.
To the purpose of fixing $Z$ at order $\als$, i.e. finding $Z_1$, 
it is sufficient to write Eq.~\eqref{renmatrix} as 
\begin{eqnarray}
 W_c^{(R)}&=&\,ZW_c+(1-Z)P_c
\notag\\
&=&\,1+\frac{C_F\als(\mu)}{rT} + \frac{4\pi C_F\als}{T} \int\frac{\d^dk}{(2\pi)^d}
\frac{e^{-i\rk}}{\k^2}\left(\frac{C_A\als}{\pi\varepsilon}+Z_1\als \right)+ \dots \,,
\label{renmatrix1}
\end{eqnarray}
where the coupling is now the renormalized strong-coupling constant in the  $\overline{\mathrm{MS}}$ 
scheme and the dots stand either for finite terms at order $\als^2$ or for terms of higher order.
Because $W_c^{(R)}$ is finite, Eq.~\eqref{renmatrix1} fixes $Z_1$; 
in the $\overline{\mathrm{MS}}$ scheme it reads 
\begin{equation}
Z_1
=-\frac{C_A}{\pi}\left(\frac{1}{\varepsilon}-\gamma_E+\ln4\pi\right)
=-\frac{C_A}{\pi}\frac{1}{\overline{\varepsilon}}\,,
\label{Z1value}
\end{equation}
where, for further use, we have defined ${1}/{\overline{\varepsilon}}\equiv {1}/{\varepsilon}-\gamma_E+\ln4\pi$.

After having removed the divergence from \eqref{renmatrix1} according to \eqref{Z1value}, 
we can perform the Fourier transform of $1/\k^2$ in three dimensions, which gives  $1/(4\pi r)$, 
and write the final expression for the renormalized cyclic Wilson loop in the $\overline{\mathrm{MS}}$ scheme:
\begin{eqnarray}
\ln W_c^{(R)} &=& \frac{C_F\als(\mu)}{rT}
\Biggl\{1+\frac{\als}{4\pi}\left[\left(\frac{31}{9}C_A-\frac{10}{9}n_f\right)
+\beta_0\left(\ln \mu^2 r^2 + 2 \gamma_E \right)\right]\Biggr.
\notag\\
&&\hspace{2.1cm}\left.+\frac{\als C_A}{\pi}
\left[1+2\gamma_E - 2\ln2 + \ln\mu^2r^2+\sum_{n=1}^\infty\frac{2(-1)^n\zeta(2n)}{n(4n^2-1)}(rT)^{2n}\right]\right\}
\notag\\
&&+\frac{4\pi\als C_F}{T}
\int\frac{\d^3k}{(2\pi)^3}
\left(  e^{i {\bf r}\cdot {\bf k}} - 1 \right) 
\left[\frac{1}{\k^2+\Pi_{00}^{(T)}(0,\k)}-\frac{1}{\k^2}\right]+C_FC_A\als^2 + {\cal O}\left(g^5\right)\,.
\notag\\
\label{renormalized}
\end{eqnarray}
The above expression is UV finite, the divergences having been reabsorbed either by the renormalization of the 
strong-coupling constant or by the subtraction of the intersection divergences along Eq.~\eqref{renmatrix}.

\subsection{The cyclic Wilson loop at short distances at leading logarithmic accuracy}
Equation \eqref{renormalized} is accurate up to next-to-leading order (NLO). 
It contains, however, logarithms in the renormalization scale $\mu$ that may be 
potentially large. These logarithms can be resummed by solving the renormalization 
group equations for $W_c^{(R)}$, which follow from the renormalization equation \eqref{renmatrix}:
\begin{equation}
\left\{
\begin{array}{c}
\displaystyle \mu \frac{\d}{\d\mu} \left(W_c^{(R)}  -P_c\right) = \gamma\, \left(W_c^{(R)}  -P_c\right)
\\
\displaystyle \mu \frac{\d}{\d\mu} \als = - \frac{\als^2}{2\pi}\beta_0
\end{array}
\right.\,.
\label{RG}
\end{equation}
The renormalized coupling and the loop functions depend on the renormalization scale $\mu$; 
the factor $\gamma$, 
\begin{equation}
\gamma \equiv \frac{1}{Z}  \mu \frac{\d}{\d\mu} Z = 2C_A\frac{\als}{\pi} + {\cal O}(\als^2)\,,
\end{equation}
is the anomalous dimension of the operator $W_c^{(R)}  -P_c$. The solution of the renormalization 
group equation is trivial and reads
\begin{equation}
\left(W_c^{(R)}  -P_c\right)(\mu) = \left(W_c^{(R)}  -P_c\right)(1/r)\;
\left( \frac{\als(\mu)}{\als(1/r)}\right)^{-4C_A/\beta_0}\,,
\label{RG1}
\end{equation}
where we have made explicit the normalization scale dependence of $W_c^{(R)}-P_c$.
Equation \eqref{RG1} implies that $\ln W_c^{(R)}$ in the $\overline{\mathrm{MS}}$ scheme 
at the scale $\mu$ may be written at NLO and leading logarithmic (LL)
accuracy (i.e. including all terms of the type $\als/(rT) \times (\als \ln\mu r )^n$) as 
\begin{eqnarray}
\ln W_c^{(R)} &=& \frac{C_F\als(1/r)}{rT}
\Biggl\{1+\frac{\als}{4\pi}\left[\left(\frac{31}{9}C_A-\frac{10}{9}n_f\right)
+ 2\, \beta_0 \gamma_E \right]\Biggr.
\notag\\
&&\hspace{2.44cm}\left.+\frac{\als C_A}{\pi}
\left[1+2\gamma_E - 2\ln2 +\sum_{n=1}^\infty\frac{2(-1)^n\zeta(2n)}{n(4n^2-1)}(rT)^{2n}\right]\right\}
\notag\\
&&+\frac{4\pi\als C_F}{T}
\int\frac{\d^3k}{(2\pi)^3}
\left(  e^{i {\bf r}\cdot {\bf k}} - 1 \right) 
\left[\frac{1}{\k^2+\Pi_{00}^{(T)}(0,\k)}-\frac{1}{\k^2}\right]+C_FC_A\als^2
\notag\\
&&
+ \frac{C_F\als}{rT}\left[\left(\frac{\als(\mu)}{\als(1/r)}\right)^{-4C_A/\beta_0} - 1 \right]
+{\cal O}\left(g^5\right)\,.
\label{renormalizedLL}
\end{eqnarray}

\section{\texorpdfstring{Renormalization up to ${\cal O}(\als^3)$}{Renormalization up to O(alpha\_s\^{}3)}}
\label{sec_als3}
The renormalization of the cyclic Wilson loop at ${\cal O}(\als^2)$
fixes the renormalization coefficient $Z_1$, and only uses the fact 
that the Polyakov-loop correlator is $1 + {\cal O}(g^3)$.
In order to provide a nontrivial check of the renormalization equation, 
we will consider now the renormalization of the cyclic Wilson loop at ${\cal O}(\als^3)$. 
At that order the expression \eqref{PL1} of the Polyakov loop correlator really matters.
We will not attempt to compute the full expression of the cyclic Wilson
loop at ${\cal O}(\als^3)$, but we will just focus on its divergent contributions
and on how they are renormalized. For this purpose, we will start with a short discussion 
about the different types of divergences and their origin. 
We will summarize here the more detailed analyses found in
Refs.~\cite{Dotsenko:1979wb} and~\cite{Brandt:1981kf}.
Then we will focus on intersection divergences.
In this section, if not otherwise stated, we will not rely on a specific gauge choice.

\subsection{Divergences}
Ultraviolet divergences come in general from integration regions in
position space where two or more vertices are contracted to one point. 
In the case of internal vertices, one gets the usual divergences removed 
through charge renormalization. But for loop functions such as the cyclic Wilson loop, 
one also gets divergences from the contraction of line vertices along the contour.
By contractions we mean only those that happen without moving vertices on a quark line to a string line or viceversa, 
and without altering the contour ordering of the vertices. 
So, for instance, in a diagram with one vertex on a quark line and two on the opposite, only the latter two can be contracted 
at the same point, either singular or smooth. For a generic diagram one has a superficial degree of divergence
\begin{equation}
\omega=1-N_{\rm ex}\,,
\label{supdiv1}
\end{equation}
at a smooth point (where the contour is differentiable and non-intersecting), and
\begin{equation}
\omega=-N_{\rm ex}\,,
\label{supdiv2}
\end{equation}
at a singular point (cusp or intersection); $N_{\rm ex}$ stands for the
number of propagators connecting the contraction point to uncontracted vertices.\footnote{
The gluon propagator satisfies the periodic boundary condition: $D_{\mu\nu}(0,{\bf x}) = D_{\mu\nu}(1/T,{\bf x})$. 
In the time interval $0\leq \tau < 1/T$ the thermal part of $D_{\mu\nu}(\tau,{\bf x})$ does not contribute to UV divergences, 
whereas these follow straightforwardly from considering the vacuum part of $D_{\mu\nu}(\tau,{\bf x})$, for instance,   
in a covariant gauge~$\xi$, 
$$
 D_{\mu\nu}(x)=\frac{\Gamma\left({D}/{2}-1\right)}{4\pi^{D/2}}\left(x^2\right)^{1-\frac{D}{2}}
\left[\frac{1+\xi}{2}\delta_{\mu\nu}+(1-\xi)\left(\frac{D}{2}-1\right)\frac{x_\mu x_\nu}{x^2}\right]\,,
$$
where $D=4-2\varepsilon$ is the number of dimensions.
The counting rules \eqref{supdiv1} and \eqref{supdiv2} apply to covariant gauges. 
They also apply to the Coulomb gauge, where, however, some diagrams that would be divergent in a covariant gauge 
vanish (e.g. the first two in the last row of Fig.~\ref{calc}).
The counting rules do not apply in general to singular gauges, where diagrams may exhibit a higher superficial degree of divergence.
Since we are dealing with gauge-invariant quantities, we are allowed to exclude the case of 
singular gauges from our considerations.
}
There are therefore three possible situations that may lead to divergences related to line vertices:
\begin{itemize}
 \item[(1)]{all vertices are contracted to a smooth point, which leads to a linear divergence;} 
 \item[(2)]{the contraction of vertices to a smooth point leaves an external propagator 
connecting a contracted to an uncontracted vertex: this leads to a logarithmic divergence that  
we will call \emph{line vertex divergence};}
 \item[(3)]{all vertices are contracted to a singular point, which gives a logarithmically 
divergent contribution; these are either \emph{cusp} or \emph{intersection divergences}.}
\end{itemize}

\begin{figure}[ht]
 \centering
 \includegraphics[width=0.6\linewidth]{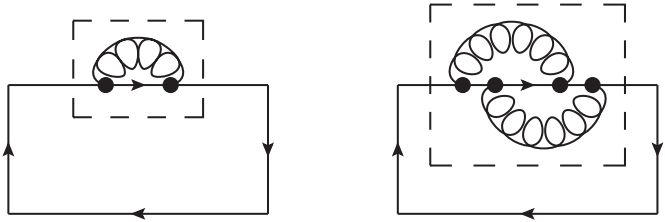}
 \caption{Examples of linear divergences. The divergences arise when the vertices inside the dashed box 
are contracted to one point.}
 \label{linear}
\end{figure}

Linear divergences are proportional to the length of the contour
and can be removed by a factor that can be interpreted as a mass term;
dimensional regularization removes these power-like divergences automatically~\cite{Polyakov:1980ca}.
Examples of diagrams with linear divergences are given in Fig.~\ref{linear}. 
In the notation adopted here, which follows the one
in Ref.~\cite{Brandt:1981kf}, the dashed box stands for integration
regions where all vertices inside the box are contracted to one point.
If the box includes a singular point, then the vertices are
contracted to that point, otherwise they can be contracted anywhere inside the box. 

\begin{figure}[ht]
 \centering
 \includegraphics[width=0.8\linewidth]{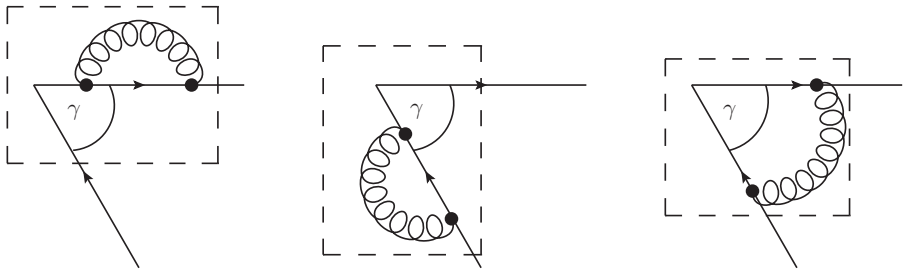}
 \caption{Contributions to a cusp divergence at $\O{}$.}
 \label{cusp}
\end{figure}

Line vertex divergences can be removed by using renormalized fields and couplings~\cite{Dotsenko:1979wb}. 
Cusp divergences arise from diagrams and integration
regions as those depicted in Fig.~\ref{cusp}. The one-loop divergence has been given 
in Eq.~\eqref{cuspdiv} as a function of the cusp angle $\gamma$. 
From it the renormalization constant for a non-cyclic Wilson loop 
(i.e. a Wilson loop with a time extension smaller than $1/T$) 
with four right-angled cusps can be inferred to be in the $\overline{\mathrm{MS}}$-scheme 
$Z=\exp\left[-{2C_F\als\mu^{-2\varepsilon}}/(\pi\bar{\varepsilon})\right]$. 
Cusp divergences are absent in a cyclic Wilson loop.

\begin{figure}[ht]
 \centering
 \includegraphics[width=0.9\linewidth]{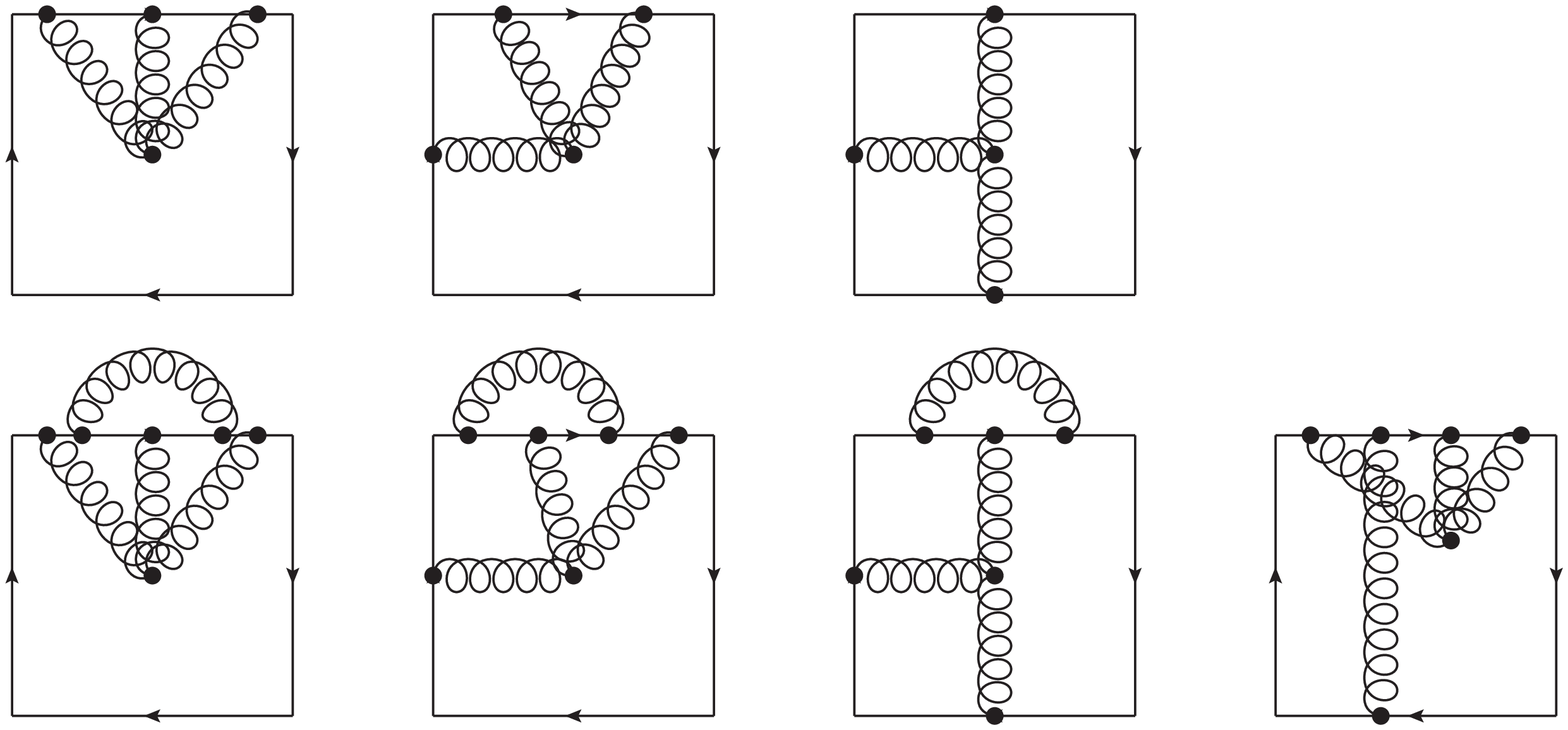}
 \caption{The diagrams in the top row are connected and do not contribute 
to intersection divergences: they either cancel through cyclicity or 
contribute to the Polyakov loop. Conversely, some of the unconnected diagrams in the bottom
row may show intersection divergences. See the text for details.}
 \label{possibilities}
\end{figure}

We turn now to the intersection divergences of the cyclic Wilson loop, 
which are our main point of interest. They only appear 
when all vertices of a diagram or subdiagram are contracted to an 
intersection point. In all cases where at least one vertex is on the string, 
if every vertex of the diagram can be contracted to the intersection,
then the contribution of the diagram cancels because of cyclicity.
If all vertices are on a quark line, then the diagram contributes 
equally to the Polyakov loop, which is finite after charge renormalization. 
This leads to the conclusion that a connected diagram cannot give rise to an intersection
divergence, because either all vertices can be contracted to an
intersection point, in which case either the divergence cancels 
because of cyclicity or because it contributes to the Polyakov loop, 
or it has at least one uncontracted vertex and therefore it is finite 
according to Eq.~\eqref{supdiv2}. These different possibilities correspond 
to the three diagrams in the first row of Fig.~\ref{possibilities}.
The first diagram has line vertices only on one quark line, so it also contributes 
to the Polyakov loop, which is finite. The second diagram has vertices on a 
string and on one quark line and thus cancels through cyclicity. 
The third diagram involves both quark lines, which means that the line
vertices cannot be contracted to the same intersection point and
therefore this diagram is finite according to \eqref{supdiv2}. 
This exhausts all possible types of connected diagrams.

\begin{figure}[ht]
 \centering
 \includegraphics[width=0.7\linewidth]{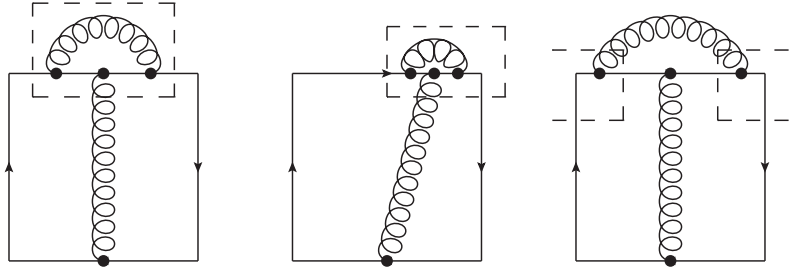}
 \caption{Intersection divergences of subdiagrams. If the line
vertices are contracted as in the left (to a smooth point) 
and middle diagram (to a singular point), we get respectively a line vertex divergence 
or a finite contribution because of the external line. 
But the periodic boundary conditions also allow for a contraction to a singular
point without external lines, as shown by the right diagram.}
 \label{subdiagram}
\end{figure}

Intersection divergences in a cyclic Wilson loop come from unconnected diagrams.\footnote{
Equation \eqref{expth} allows us to restrict to 2PI unconnected diagrams.}
In particular they come from unconnected diagrams made of at least one 
subdiagram with vertices on both quark lines and a subdiagram 
that is divergent once all its vertices are contracted to an intersection point.
This can be understood by looking at the Feynman diagrams in the second row of Fig.~\ref{possibilities}. 
The first diagram is part of the Polyakov loop, which is finite,  
the second diagram cancels because of cyclicity, but the third
and fourth diagrams are divergent, because we can contract the line
vertices of the respective one-gluon and three-gluon subdiagrams to
an intersection point. The periodic boundary conditions are of crucial
importance here, because in order to have no external lines, the
intersection point must be approached from the left and the right, as
depicted in Fig.~\ref{subdiagram}.

As we just argued, diagrams contributing to intersection divergences
need to contain at least one subdiagram with vertices on both quark
lines. We call such subdiagrams \emph{bases}. To be more specific, for
each intersection-divergent diagram we define its basis as the
subdiagram which is of highest order in $\alpha_s$ and still
finite. In other words, the basis of a diagram is the subdiagram which
is obtained by removing all parts that contribute to the divergence,
but no more than that. Conversely, we can obtain all
intersection-divergent diagrams by considering all possible bases and
combining them with all subdiagrams that lead to an intersection
divergence. This classification of the intersection-divergent diagrams
according to their bases and intersection-divergent subdiagrams
provides the framework for the following analysis.

\begin{figure}[ht]
 \centering
 \includegraphics[width=0.9\linewidth]{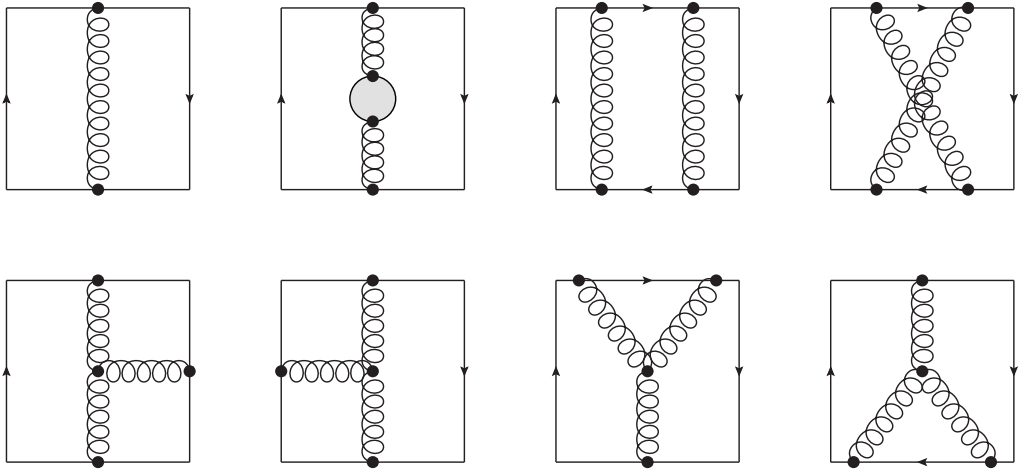}
 \caption{Bases relevant for the intersection divergences of $\O{3}$.}
 \label{bases}
\end{figure}

All bases relevant for the computation of 
intersection divergences at $\O{3}$ are given by the eight diagrams shown in Fig.~\ref{bases}.
Note that bases may be reducible, provided that adding intersection-divergent subdiagrams makes
them 2PI.
In order to get all diagrams contributing to intersection
divergences at this order, we need to add subdiagrams of $\O{2}$ to the first basis, and
a single gluon exchange to the other bases, in such a way that their line
vertices can be contracted to at least one of the intersection points.

\begin{figure}[ht]
 \centering
 \includegraphics[width=\linewidth]{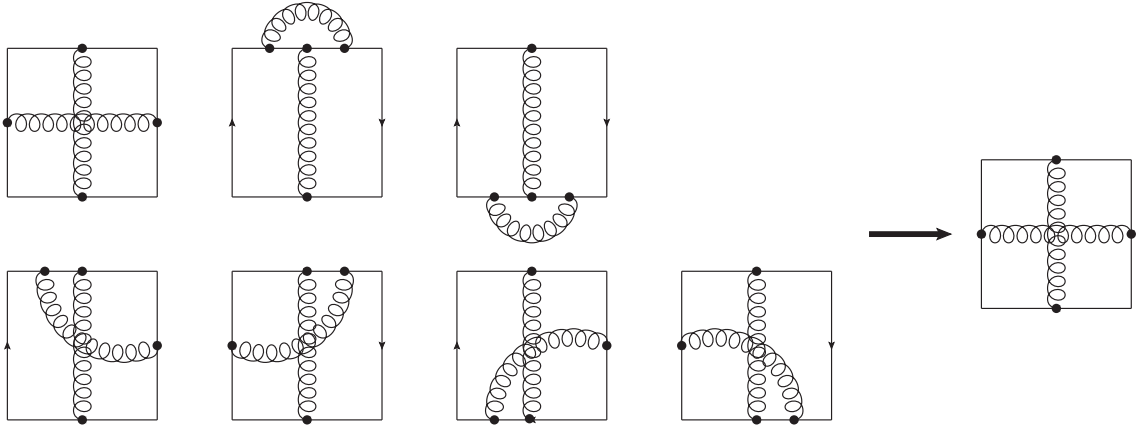}
 \caption{A class of 2PI diagrams represented by the diagram on the right. 
In Coulomb gauge the diagram on the right is the only one that does not vanish, see Sec.~\ref{sec_pert}.
}
 \label{abbr}
\end{figure}

The obtained diagrams may be divided into \emph{classes}.
A class of diagrams is made of diagrams that have the same basis and 
that can be transformed one into the other by just moving some of the line vertices across 
the intersection point from string to quark line or from quark line to string without 
changing their ordering along the contour of the Wilson loop. 
A class of $\O{2}$ diagrams is shown in Fig.~\ref{abbr}; a class 
will be typically represented by just one of its diagrams.

\subsection{Renormalization}
\begin{figure}[ht]
 \centering
 \includegraphics[width=0.25\linewidth]{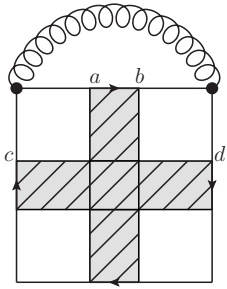}
 \caption{Class of diagrams with a one-gluon subdiagram.}
 \label{abcd}
\end{figure}

We consider first the effect of a single gluon when added to one of the diagrams in Fig.~\ref{bases}. 
It generates two classes of diagrams that lead to intersection divergences:
one is shown in Fig.~\ref{abcd} with an intersection divergence at $\r/2$, 
the other one would have the gluon attached to the lower quark like and an intersection divergence at $-\r/2$.
In Fig.~\ref{abcd}, the contour integration for the left vertex of the gluon goes from
$(0,c\r)$ to $(a/T,\r/2)$, and the contour integration for the right vertex 
goes from $(-(1-b)/T,\r/2)$ to $(0,d\r)$, where we have used the
periodic boundary conditions to shift the right vertex by $-1/T$ in time; 
the grey area stands for a generic basis. 
Colour factors are considered separately. In a generic covariant gauge, the result reads:
\begin{align}
 -\frac{\als}{\pi}&\left[\frac{1}{\bar{\varepsilon}}+2\gamma_E-\ln4+2
+\ln\frac{\mu\,a(1-b)}{T(1-b+a)}+\ln\frac{\left(\frac{1}{2}-c\right)\left(\frac{1}{2}-d\right)}{|c-d|}r\mu\right.
\notag\\
 &\Biggl.\hspace{5pt}+\frac{1}{2}(1-\xi)\ln\frac{\sqrt{\left(\frac{1}{2}-d\right)^2r^2T^2+a^2}
\sqrt{\left(\frac{1}{2}-c\right)^2r^2T^2+(1-b)^2}}{(1-b+a)\,|c-d|\, rT}\hspace{5pt}\Biggr]\,.
\label{crossdiag}
\end{align}
The divergent part does not depend on the contour parameters $a$,
$b$, $c$, $d$, hence, it can be factored out from the contour integration. 
Because the contour parameters only appear in logarithms, they do not introduce new divergences when the
contour integration for the basis is performed, since the basis is by definition free from intersection divergences.
The fact that the divergence factorizes can be intuitively understood by noting 
that it comes from integration regions where all vertices of the added subdiagram are at the
intersection, so it is unaffected by the contour integration for the basis (see also Ref.~\cite{Brandt:1981kf}). 
Note that the divergent part of \eqref{crossdiag}, i.e. $-\als/\pi \times 1/\overline{\varepsilon}$,
does not depend on the gauge fixing parameter~$\xi$; 
furthermore, the same divergent part occurs also in Coulomb gauge, 
in which case, however, only the integration along the strings contributes.
The second class of diagrams with the gluon attached to the 
lower quark line gives the same divergence as in \eqref{crossdiag}.

At order $\O{2}$ the only class of divergent diagrams is shown in Fig.~\ref{abbr}.
The divergence of that class is $\widetilde{C}_2 \times 2 \times (-\als/\pi \times 1/\overline{\varepsilon}) \times $
(basis, i.e. the one-gluon exchange diagram, $\als/r$); $\widetilde{C}_2$ is the colour-connected 
coefficient of the diagrams, the factor 2 comes from the two 
intersection divergences at ${\r}/{2}$ and  $-{\r}/{2}$, and the divergence 
is the one calculated in \eqref{crossdiag}. The cancellation of this
divergence in the $\overline{\mathrm{MS}}$-scheme fixes $Z_1$ to the value calculated 
in \eqref{Z1value}.

\begin{figure}[ht]
 \centering
 \includegraphics[width=0.5\linewidth]{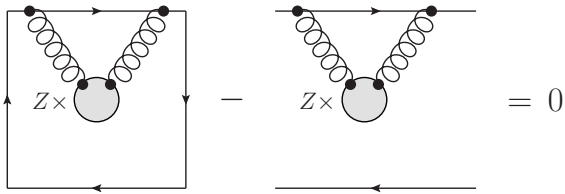}
 \caption{Cancellation of contributions from $W_c$ and $P_c$. The Polyakov-loop correlator is represented without strings.}
 \label{loopZ1a}
\end{figure}

Both $W_c$ and $P_c$ are equal to $1$ at zeroth order in $\als$.
At that order the renormalization condition is automatically fulfilled: $Z+(1-Z)=1$.
A similar cancellation happens at any order (including odd powers of $g$ 
from contributions of the Debye mass scale) when the renormalization constant 
multiplies diagrams that occur identically in both the cyclic Wilson loop 
and the Polyakov loop correlator. These are contributions to the cyclic Wilson loop 
coming from the gluon self-energy diagrams in the top row of Fig.~\ref{calc}
(second and third diagram) and from the
diagrams in the bottom row of Fig.~\ref{calc} (third and fourth diagram), 
and the contribution to the Polyakov loop correlator coming from 
the Polyakov loop, which is $P_L^2-1 \approx 2\,\delta P_L$ in Eq.~\eqref{PL1}.
An example of cancellation is shown in Fig.~\ref{loopZ1a}.

\begin{figure}[ht]
 \centering
 \includegraphics[width=0.5\linewidth]{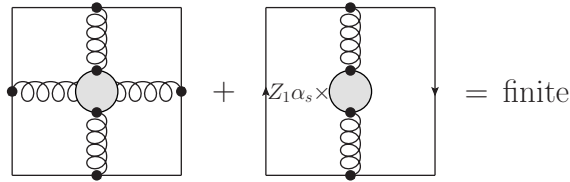}
 \caption{The divergences of the two classes of diagrams cancel for the same reason as 
 the lower-order classes of diagrams without self-energy insertions do.
 The gluon connecting the strings does not interact with the bubble.}
 \label{loopZ1b}
\end{figure}

Let us discuss now divergences coming from $\O{3}$ unconnected diagrams in 
the cyclic Wilson loop. We start by considering classes of diagrams whose basis is the second 
diagram in the first row  of Fig.~\ref{bases}.
The cancellation of the divergences carried by these diagrams 
in the renormalization equation is a simple extension of the 
cancellation that happens at $\O{2}$ between the intersection divergence 
carried by the last diagram of the second row of Fig.~\ref{calc} and  $Z_1\als$ times  
the one-gluon exchange diagram. This is illustrated in Fig.~\ref{loopZ1b}, where the bubble stands 
for any self-energy insertion. Also in this case, odd powers of $g$ coming from the Debye mass scale, 
such as a term arising at order $g^5$ in our adopted hierarchy of energy scales, cancel out.

\begin{figure}[ht]
 \centering
 \includegraphics[width=\linewidth]{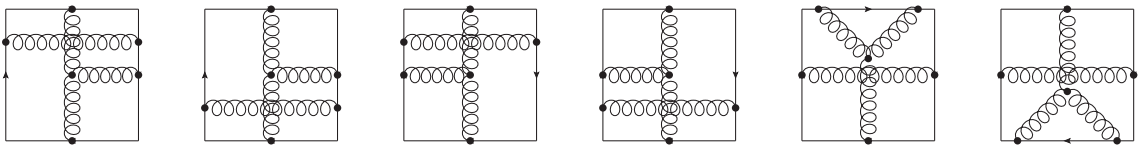}
 \caption{Classes of diagrams with an intersection divergence and a three-gluon diagram as basis.}
 \label{3Gdiv}
\end{figure}

The cancellation of divergences carried by classes of diagrams whose bases are from the second
row of Fig.~\ref{bases} (shown in Fig.~\ref{3Gdiv}), 
which we will call collectively $\gamma_T$, is similar to the one discussed 
in the previous paragraph. They are cancelled by $Z_1\als$ times the respective
basis diagram from the $\O{2}$ expansion of $W_c$. As shown by Eq.~\eqref{crossdiag} and 
the following discussion, adding one gluon to a basis  
gives a divergent factor, which is $2 \times \left(-{\als}/({\pi\bar{\varepsilon}})\right)$. 
The colour factor of the bases is $iC_FC_A/2$, the colour-connected coefficient
of diagrams like those in Fig.~\ref{3Gdiv} is $- i C_FC_A^2/4$. With this it is
easy to see that the divergence of a class of diagrams with basis $\gamma_T$ is:
\begin{equation}
 \left(-\frac{2\als}{\pi\bar{\varepsilon}}\right)\left(-\frac{i}{4}C_FC_A^2\right)W\left(\gamma_T\right)=
-Z_1\als\left(\frac{i}{2}C_FC_A\right)W\left(\gamma_T\right)\,,
\end{equation}
where $W\left(\gamma_T\right)$ denotes the value of a diagram 
belonging to $\gamma_T$ without its colour factor. We see that all these
divergences cancel in $ZW_c$ at $\O{3}$.

\begin{figure}[ht]
 \centering
 \includegraphics[width=\linewidth]{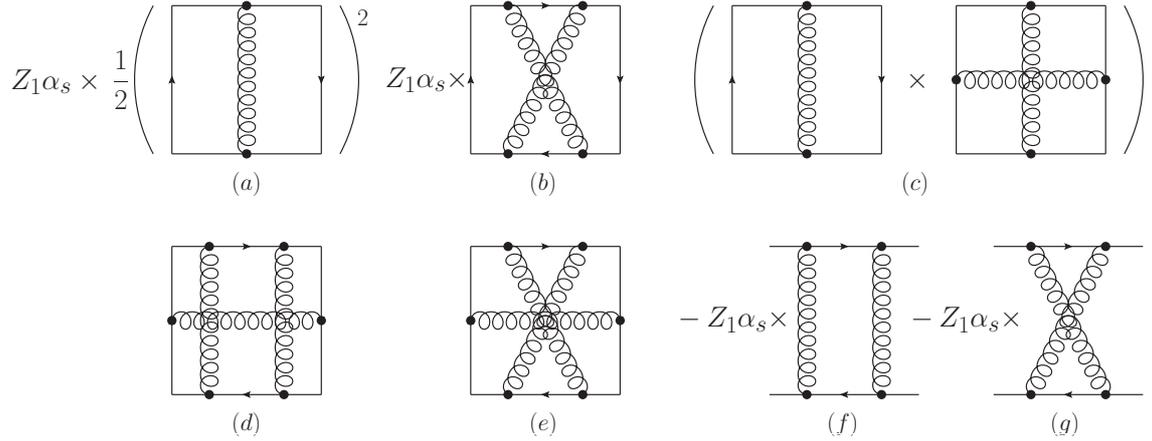}
 \caption{The sum of all these contributions is finite. It includes
   all diagrams with bases made of two gluons exchanged between the
   quark lines, and related contributions from both the cyclic Wilson
   loop and the Polyakov loop correlator (diagrams $(f)$ and $(g)$).}
 \label{cancel}
\end{figure}

With the cancellation of divergences associated to classes of diagrams 
whose basis is either the third or fourth diagram in the first row 
of Fig.~\ref{bases}, we come to the first non-trivial check of the renormalization equation \eqref{renmatrix}. 
We call the basis with two ladder gluons (i.e. the third diagram in the first row  of Fig.~\ref{bases}), $\gamma_{II}$, 
and the one with two crossed gluons (i.e. the fourth diagram in the first row  of Fig.~\ref{bases}), $\gamma_X$. 
Without colour factors the sum of these two bases
gives just half of the square of the  ${\cal O}(\als)$ diagram, 
so $W\left(\gamma_{II}\right)+W\left(\gamma_X\right)={\als^2}/({2r^2T^2})$. 
The colour-connected coefficients for the diagrams with $\gamma_{II}$ and $\gamma_X$ as bases 
(these are the diagrams $(d)$ and $(e)$ of Fig.~\ref{cancel})
are $C_FC_A^2/4$ and $C_FC_A^2/2$ respectively.
The divergences associated to these classes of diagrams are of the type of Fig.~\ref{abcd} 
and bring a factor $-\als/\pi \times 1/\bar{\varepsilon}$ for each intersection point.
In order to cancel these divergences, we have to collect also divergences coming from 
$Z_1\als \times W_c$ that are not of the type cancelled by the Polyakov loop  
and divergences coming from the expansion of the exponential on the right-hand side of Eq.~\eqref{expth}
(recall that the exponent is made by 2PI diagrams only). 
We will highlight only the divergent terms and use the symbol ``$\supset$'' with the meaning 
``includes the divergent term''.  We have (compare with Fig.~\ref{cancel}):
\begin{itemize}
\item[$(a)$]{the renormalization constant $Z_1\als$ multiplying the
second order expansion term of the one-gluon exchange diagram:
$\displaystyle Z_1\als \times W_c\supset
\left(-\frac{C_A\als}{\pi\bar{\varepsilon}}\right) \times \frac{1}{2} \left(\frac{C_F\als}{rT}\right)^{2}$;}
\item[$(b)$]{the renormalization constant $Z_1\als$ multiplying the basis $\gamma_X$
(whose colour-connected coefficient is $\widetilde{C}_2 = -C_FC_A/2$):
$\displaystyle Z_1\als \times W_c\supset
\left(-\frac{C_A\als}{\pi\bar{\varepsilon}}\right) \times \left(-\frac{C_FC_A}{2}W(\gamma_{X})\right)$;}
\item[$(c)$]{the one-gluon exchange diagram times the intersection-divergent
diagrams at $\O{2}$ from the expansion of the exponential:
$\displaystyle W_c\supset\left(\frac{C_F\als}{rT}\right) \times 
\left(\frac{C_A\als}{\pi\bar{\varepsilon}}\right)\left(\frac{C_F\als}{rT}\right)$;}
\item[$(d)+(e)$]{the divergent contributions from diagrams with bases $\gamma_{II}$ and $\gamma_X$:\\
$\displaystyle W_c\supset
\left(-\frac{2\als}{\pi\bar{\varepsilon}}\right)\left[\frac{1}{4}C_FC_A^2W(\gamma_{II})+\frac{1}{2}C_FC_A^2W(\gamma_X)\right]$.}
\end{itemize}
Summing up these contributions, we get
\begin{align}
 ZW_c\supset&\left(-\frac{C_FC_A^2\als}{2\pi\bar{\varepsilon}}\right)\left[W(\gamma_{II})+2W(\gamma_X)-W(\gamma_X)\right]
-\frac{C_F^2\als^2}{2r^2T^2}\frac{C_A\als}{\pi\bar{\varepsilon}}+\frac{C_F^2\als^2}{r^2T^2}\frac{C_A\als}{\pi\bar{\varepsilon}}
\notag\\
=&\left(C_F^2-\frac{1}{2}C_FC_A\right)\frac{\als^2}{2r^2T^2}\frac{C_A\als}{\pi\bar{\varepsilon}}\,.
\end{align}
This is exactly cancelled by the remaining contributions coming from the Polyakov loop correlator \eqref{PL1}, 
given as $(f)+(g)$ in Fig.~\ref{cancel}:
\begin{equation}
(1-Z)P_c\supset \frac{C_A\als}{\pi\bar{\varepsilon}}\times \left[-\left(C_F^2-\frac{1}{2}C_FC_A\right)\frac{\als^2}{2r^2T^2}\right]\,.
\end{equation}
The cancellation provides a non-trivial verification of the renormalization equation  \eqref{renmatrix}.
Note that the cancellation of the diagrams in Fig.~\ref{cancel} is gauge independent, since 
$Z_1$, the one-gluon exchange diagram, the Polyakov-loop correlator as well as the 
combination  $W\left(\gamma_{II}\right)+W\left(\gamma_X\right)$ are gauge invariant.

\begin{figure}[ht]
 \centering
 \includegraphics[width=\linewidth]{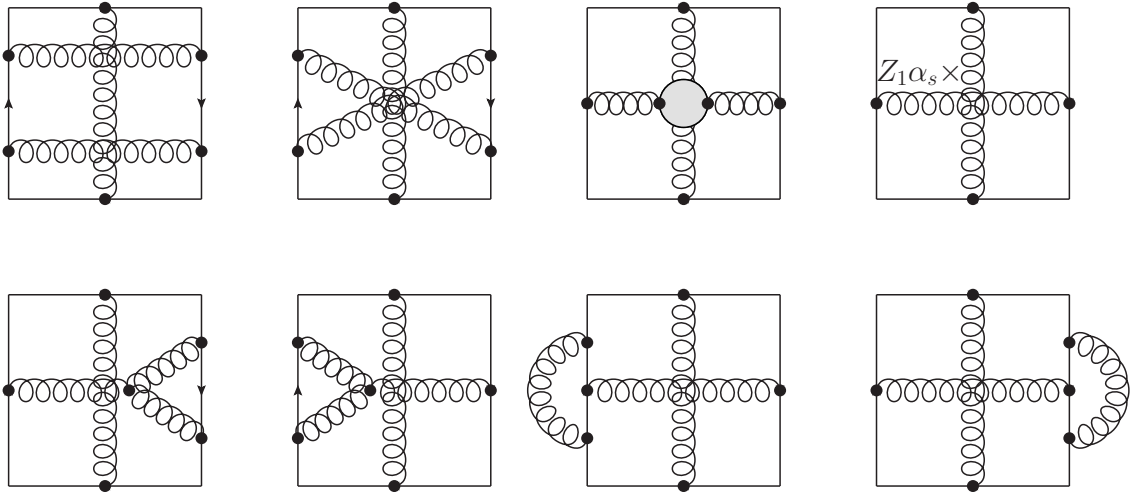}
 \caption{Classes of diagrams contributing to $Z_2$.}
 \label{Z2determ}
\end{figure}

Finally, we observe that the second-order expansion term of the renormalization constant 
times ${\cal O}(\als)$ diagrams of the cyclic Wilson loop gives a divergent term like
\begin{equation}
ZW_c\supset Z_2\als^2 \times \frac{C_F\als}{rT}\,,
\label{Z2}
\end{equation}
where $Z_2$ may contain single or double poles in $1/\varepsilon$.
This term is necessary to cancel divergent contributions coming from unconnected diagrams that have 
the one-gluon exchange diagram, i.e. the first diagram of Fig.~\ref{bases}, as a basis and subdiagrams 
of ${\cal O}(\als^2)$ added. The different diagram classes are shown in Fig.~\ref{Z2determ}. 
The sum of all divergent terms of these diagrams (after charge renormalization) determines the value of~$Z_2$.

Because the right-hand side of Eq.~\eqref{Z2} is proportional to $1/(rT)$, 
in order to complete our proof of renormalizability at $\O{3}$ 
we have to make sure that the sum of diagrams in Fig.~\ref{Z2determ} only contains 
divergent terms proportional to $1/(rT)$. The second and the third classes of diagrams in 
the first row and the classes of diagrams in the second row of Fig.~\ref{Z2determ}
do indeed fulfill this criterion. The reason is that, in order to get an intersection divergence from these diagrams, 
we have to contract all gluons but the one connecting the two quark lines to the intersection points,
otherwise there would be a gluon propagator connecting the intersection point 
with an uncontracted vertex and from Eq.~\eqref{supdiv2} we know that such a configuration 
is convergent. As a consequence, since the divergence at the contraction does not depend on $r$, the complete $r$ 
dependence comes from the one-gluon exchange, which is proportional to $1/(rT)$.
The same reasoning applies to the first and fourth classes of diagrams in the first row 
if we contract all gluons but the one connecting the two quark lines to the intersection points.
Such kind of intersection divergence is called \emph{overall divergence} in~\cite{Brandt:1981kf}, 
to distinguish it from a \emph{subdivergence}, which is a divergence that 
occurs when only one of the gluons is contracted to the intersection points.
Consider the first diagram in Fig.~\ref{Z2determ}. Its colour-connected coefficient is 
$C_FC_A^2/4$ just like diagram $(d)$ of Fig.~\ref{cancel}. A~subdivergence may come 
either by contracting the upper gluon connecting the strings to the intersection point at $\r/2$ 
and keeping the finite part of the remaining diagram, or by contracting the lower gluon 
to the intersection point at $-\r/2$ and keeping again the finite part of the remaining diagram.
The divergence may be read from \eqref{crossdiag} and is $-\als/(\pi\bar{\varepsilon})$; the 
finite part is the same for the two situations and we call it $W(\gamma_{+})^{\rm (finite)}$.
Hence, the subdivergence of the first diagram in Fig.~\ref{Z2determ} is 
$\displaystyle \frac{1}{4}C_FC_A^2\left(-\frac{2\als}{\pi\bar{\varepsilon}}\right) W(\gamma_{+})^{\rm (finite)}$.
Because $W(\gamma_{+})^{\rm (finite)}$ has a more complicate functional dependence than just $1/(rT)$,  
this is a divergence that is not canceled by the right-hand side of \eqref{Z2}.
Consider now the fourth diagram in Fig.~\ref{Z2determ} when none of the gluons is contracted 
to an intersection point. The colour-connected coefficient is $\widetilde{C}_2 = -C_FC_A/2$.
The divergence is $\displaystyle (Z_1\als) \times \left(-\frac{1}{2}C_FC_A \right)\, W(\gamma_{+})^{\rm (finite)}$.
Note that the finite part is the same multiplying the subdivergence in the first diagram.
Summing up the (sub)divergences of the two diagrams we obtain
\begin{equation}
ZW_c\supset \frac{1}{4}C_FC_A^2\left(-\frac{2\als}{\pi\bar{\varepsilon}}\right)
W(\gamma_{+})^{\rm (finite)}+ \left(-\frac{C_A\als}{\pi\bar{\varepsilon}}\right) \left(-\frac{1}{2}C_FC_A\right)\,W(\gamma_{+})^{\rm (finite)}=0\,,
\end{equation}
where we have used the value of $Z_1$ derived in \eqref{Z1value}.
This completes our proof of renormalizability of the intersection divergences at ${\cal O}(\als^3)$.

\section{The cyclic Wilson loop at large distances}
\label{sec_largedistance}
Equation \eqref{renmatrix} establishes how the cyclic Wilson loop renormalizes, i.e. 
by mixing with the Polyakov loop correlator. Although we have tested the equation in 
a specific hierarchy of energy scales, i.e. $1/r\gg T\gg m_D \gg \als/r$, its validity is 
not bound to this hierarchy, as we anticipated in Sec.~\ref{sec_pert}, for it follows from general arguments based ultimately only on the 
UV behaviour of QCD and on the geometry of the loop functions~\cite{Brandt:1981kf}. In particular, Eq.~\eqref{renmatrix}
should also hold for cyclic Wilson loops at large distances, i.e. for $r m_D \sim 1$.

Let us consider in the long-distance case $r m_D \sim 1$ the UV divergences of the cyclic Wilson loop 
at ${\cal O}(\als^2)$. To make contact with Sec.~\ref{sec_pert} we work in Coulomb gauge.
The divergent diagram is again the last diagram 
in the second row of Fig.~\ref{calc}. The divergence comes from the vacuum part of the transverse 
gluon connecting the two strings. As expected, this quantity is insensitive to the low energy dynamics:
the thermal part of the transverse gluon gives a finite contribution, the temperature providing 
a UV cut-off through the Bose--Einstein distribution, moreover transverse gluons are not screened.
The temporal gluon connecting the quark lines is instead screened 
by the Debye mass. This is the only difference with the calculation performed in Sec.~\ref{sec_pert}.
Because the temporal gluon factorizes, at long distances the leading divergence of the cyclic Wilson loop 
due to intersection may be easily inferred from Eq.~\eqref{uvdiv}. 
Adding to it the leading-order diagram, which consists of the first diagram in Fig.~\ref{tree}, 
also with a screened temporal gluon, we get 
\begin{eqnarray}
W_c &=&\,1+ \frac{4\pi C_F\als(\mu)}{T}\frac{e^{-m_Dr}}{4\pi r} 
+ \frac{4 C_FC_A\als^2}{T} \,\frac{e^{-m_Dr}}{4\pi r}\,\frac{1}{\varepsilon} + \dots \,,
\label{renmatrix2}
\end{eqnarray}
where $\exp(-m_Dr)/(4\pi r)$ is the Fourier transform in three dimensions of the screened temporal 
gluon propagator, $D_{00}(0,{\bf k})=1/(\k^2+m_D^2)$, and the dots stand either for finite terms or for terms of higher order.
It is straightforward to see that the expression in \eqref{renmatrix2} is renormalized by \eqref{renmatrix} with 
the same renormalization constant $Z$ computed in \eqref{zexpand} and \eqref{Z1value}.
This confirms, at least at leading order, our expectation that Eq.~\eqref{renmatrix}
renormalizes the cyclic Wilson loop at any distance. Finally, we note that our conclusion 
and in particular Eq.~\eqref{renmatrix2} disagree with the long-distance finding of~\cite{Burnier:2009bk}.\footnote{
The disagreement may be traced back to the contribution from the non-zero modes to the transverse gluon 
connecting the strings in the last diagram in the second row of Fig.~\ref{calc}.
In our computation, the relevant integral reads 
\begin{eqnarray*}
\sum_{n\neq 0} \int \frac{\d^dq}{(2\pi)^d}\frac{\left(e^{i {\bf r}\cdot {\bf q}} + e^{-i {\bf r}\cdot {\bf q}} - 2\right) r^2}
{2 (\r\cdot\q)^2\left(\q^2 + \omega_n^2\right)}
&=&
\frac{1}{8\pi^2 T}\left[\frac{1}{\varepsilon} + \gamma_E - \ln 4\pi + \ln \frac{\mu^2}{T^2} 
+  2 \int_1^\infty \frac{\d x}{x^2} \, \ln \left(1-e^{-2\pi r T x}\right)
\right. 
\\
&& \hspace{1cm}
\left.
+ \pi r T\left(
\frac{1}{\varepsilon} - \gamma_E + \ln 4\pi + \ln \frac{\mu^2}{T^2} 
\right)
\right]\,.
\end{eqnarray*}
The first line agrees with Eq.~(A.19) of \cite{Burnier:2009bk}, but the second line, which comes 
from the non-vanishing contributions from the double pole at $\r\cdot\q =0$ in the two half-planes where the two Fourier
exponentials separately converge, is missing. 
It is precisely the $1/\varepsilon$ singularity in the second line that cancels the UV divergence coming 
from the zero modes, while the singularity in the first line, combined with other contributions, 
leads eventually to the result~\eqref{renmatrix2}.
We thank the authors of~\cite{Burnier:2009bk} for communication on this point. 
}

\section{Conclusions}
\label{sec_concl}
We have investigated the UV behaviour of the cyclic Wilson loop. 
Our short-distance, order-$g^4$ calculation, summarized in Eq.~\eqref{finalunrenormalized}, 
confirms the finding of Ref.~\cite{Burnier:2009bk}, namely that the cyclic Wilson loop is divergent
after charge renormalization and that its divergence is not a standard cusp divergence.

In Sec.~\ref{sec_renorm}, we have shown how the periodic boundary conditions 
influence the divergences and the renormalization properties of the cyclic Wilson loop.
The contour can be seen as having intersection points along the two strings, but  
only the two endpoints are relevant for renormalization purposes. 
Applying the intersection-divergence renormalization technique of Ref.~\cite{Brandt:1981kf},  
we have obtained the renormalization equation for the cyclic Wilson loop:  $W^{(R)}_c=Z W_c+(1-Z)P_c$.
This equation represents the main result of the paper and shows how the cyclic Wilson loop, $W_c$, 
mixes under renormalization with the correlator of two Polyakov loops, $P_c$, their mixing being determined by the
renormalization constant~$Z$. The equation holds for all $SU(N_c)$ gauge theories; 
in the case of an abelian gauge theory, the cyclic Wilson loop and the correlator of Polyakov loops coincide 
and are finite. 

We have determined the renormalization constant, $Z$, up to order $g^2$, obtaining 
$Z=1- {C_A\als\mu^{-2\varepsilon}}/{\pi} \times \left({1}/{\varepsilon}-\gamma_E+\ln4\pi \right)$ 
in the $\overline{\mathrm{MS}}$ scheme. In Sec.~\ref{sec_als3}, we have verified 
that this expression of $Z$ reabsorbs all divergences of the type $\als^3/(rT)^2$; 
this is a non-trivial check involving the Polyakov-loop correlator at order $\als^2$.
From the renormalization constant we could extract the intersection anomalous dimension 
at one loop and solve the corresponding renormalization group equations \eqref{RG}.
The result provides the cyclic Wilson loop at LL accuracy \eqref{renormalizedLL}, which is the 
novel computational outcome of this work. 

Finally, we observe that the renormalization condition \eqref{renmatrix} is equivalent 
to stating that the combination $W_c-P_c$ is multiplicatively renormalizable.\footnote{
In dimensional regularization this implies that the ratio $(W_c-P_c)(r)/(W_c-P_c)(r_0)$,
where the loop functions in $(W_c-P_c)(r)$ are evaluated at a distance $r$ while 
those in $(W_c-P_c)(r_0)$  are evaluated at a fixed distance $r_0$, is finite. 
On the lattice, because of linearly divergent renormalization factors proportional to $r$ and $1/T$, 
a possible finite quantity is $(W_c-P_c)(r)/(W_c-P_c)(r_0) \times (W_c-P_c)(2r_0-r)/(W_c-P_c)(r_0)$.
} 
The combination $W_c-P_c$ is therefore an ideal quantity to be computed on the lattice, while 
clearly the cyclic Wilson loop is not (see discussion in~\cite{Burnier:2009bk}).
It could provide a new, independent and gauge-invariant lattice observable for the study 
of the thermodynamical properties of two static sources in a thermal bath, relevant for quarkonium 
physics in a quark-gluon plasma, being at the same time well suited for comparisons with analytic studies 
like the one performed in this work.

\section*{Note added}
After this paper was completed we became aware of Ref.~\cite{Korchemskaya:1994qp},
which analyzes the intersection divergence of two straight Wilson lines crossing at a point in Minkowski space. The resulting anomalous dimension
matrix is provided up to order $\als^2$ as a function of the Minkowskian angle $\gamma$ between the lines.
We find that, after relating their path $(W_1)^{ij}_{ij}$ to $N_c^2 P_c$
and $(W_2)^{ij}_{ij}$ to $N_c W_c$, the order-$\als$ result in Eq.~(3.21) of \cite{Korchemskaya:1994qp}
is consistent with ours for $\gamma=i\pi/2$.\footnote{
One has to consider that the anomalous dimension in Eq.~(2.10) of~\cite{Korchemskaya:1994qp}
refers to a single intersection point, while in the cyclic Wilson loop case one needs to consider two.
Furthermore, our geometry at the intersection differs from theirs not only in the signature
of spacetime but also in the fact that in our case the string is not crossing the intersection point, but coming back onto itself. However this does not affect the comparison,
as we can change our geometry to the Euclidean version of theirs by changing
$\gamma\to i\pi-\gamma$ for one of the two strings at the intersection, which,
for $\gamma=i\pi/2$, leaves the angle unchanged. Indeed, the authors of \cite{Korchemskaya:1994qp}
make wide use of the $\gamma\to i\pi-\gamma$ symmetry in their calculation.}
Moreover their intersection anomalous dimension at order $\als^2$, for $\gamma=i\pi/2$,
turns out to be completely determined by the cusp anomalous
dimension at the same angle, which is in agreement with a preliminary calculation of ours. We plan to return to this topic elsewhere.

\acknowledgments

We acknowledge financial support from the DFG cluster of excellence
\emph{Origin and structure of the universe}
(www.universe-cluster.de). This research is supported by the DFG
grant BR 4058/1-1. The work of J.G. was supported by the Natural
Science and Engineering Research Council of Canada and by an Institute
of Particle Physics Theory Fellowship. Part of the work of M.B. was
done at Kyoto University and supported by the Global Cluster of
Excellence (GCOE) in the framework of the Bilateral International
Exchange Program (BIEP) of 2012; M.B. thanks Prof.~Hideo Suganuma 
and his group for their warm hospitality.

\end{document}